\documentclass[aps,prb,prX,citeautoscript,a4paper,twocolumn]{revtex4}

\pdfoutput=1
\usepackage{graphicx}
\usepackage{amsmath}
\usepackage{amssymb}
\begin{document}
 \title{Correlation-induced corrections to the band structure of boron nitride: a wave-function-based approach}

\author{A.\ Stoyanova}
\email[Corresponding authtor:]{alex07@mpipks-dresden.mpg.de}
\author{L.\ Hozoi}
\author{P.\ Fulde}
\affiliation{Max-Planck-Institut f\"ur Physik komplexer Systeme, \\
                   N\"othnitzer Strasse 38, 01187 Dresden, Germany
                   }
\author{H. Stoll}
\affiliation{Universit\"at Stuttgart, Pfaffenwaldring 57, 70550 Stuttgart, Germany                    
 }

% ABSTRACT
\begin{abstract}
We present a systematic study of the correlation-induced corrections to the electronic band structure of zinc-blende BN. Our investigation employs an \textit{ab initio} wave-function-based local Hamiltonian formalism which offers a
rigorous approach to the calculation of the polarization and local charge redistribution effects around an extra electron or hole placed into the conduction or valence bands of semiconducting and insulating materials. Moreover, electron correlations beyond relaxation and polarization can be readily incorporated. The electron correlation treatment is performed on finite clusters. In conducting our study, we make use of localized Wannier functions and embedding potentials derived explicitly from prior periodic Hartree-Fock calculations. The on-site and nearest-neighbor charge relaxation bring corrections of several eV to the Hartree-Fock band gap. Additional corrections are caused by long-range polarization effects. In contrast, the dispersion of the Hartree-Fock bands is marginally affected by electron correlations.
Our final result for the fundamental gap of zinc-blende BN compares well with that derived from soft x-ray experiments at the B and N $K$-edges. 
\end{abstract}
\maketitle

\section{Introduction}
Electronic structure calculations for solids are commonly based on solving self-consistently a single-particle Schr\"odinger equation. The solutions have the form of Bloch waves, i.e., they are translation symmetry adapted. First principles calculations of this form are successfully performed at present within density functional theory (DFT) \cite{Hoh, KS} or Hartree-Fock (HF)\cite{HF} approaches supplemented by post-HF calculations. A particular object of interest in these studies is the correct treatment of electron interactions in crystals with relatively weak as well as strong electron correlations.  

Within the framework of DFT electron correlation effects are incorporated in practice through various exchange-correlation functionals, whereas diverse post-HF correlation algorithms have been designed to address this issue with the wave-function-based approach \cite{Sun, Ayala, Pisani, Stoll1, Gr2, Shukla2, GF3, FuldeAdv, Birken2, Forner, Karin, Hirata2, Hozoi, Albrecht}. DFT is strictly speaking a ground-state theory. Nevertheless the eigenvalues of the Kohn-Sham (KS)  equations have been interpreted as the electronic energy bands. 
Recent theoretical and computational studies on molecular systems have shown that in analogy with Koopmans' theorem in HF theory, the absolute values of the energies of the KS (spin-) orbitals can be indeed interpreted as approximate vertical ionization potentials \cite{Gritsenko1, Gritsenko2}.
For the highest occupied KS orbital, this relation becomes exact \cite{Janak, PerdewIP, Gritsenko1, Gritsenko2}. Yet, the calculated ionization potentials depend on the approximation to the exact KS potential.    

The description of the electronic properties of many elemental metals and various covalent and ionic compounds by means of DFT has appeared to be very successful. Recent extensions of the theory to the time-dependent domain, time-dependent DFT and current DFT \cite{GrossRunge, TDDFT, TDCDFT}, have also allowed for a more rigorous treatment of optical excitations. 
By circumventing the calculation of the many-body wave function, ground or excited state, DFT offers a relatively simple approach to the electronic structure of solids. Such studies are commonly carried out within the local-density approximation (LDA) \cite{Perdew} or generalized gradient approximation (GGA) \cite{Perdew2}.  

It is well known, however, that the LDA-DFT and GGA-DFT based calculations underestimate the band gaps of insulators and semiconductors. It is also known, that this failure is only partially related to the use of approximate exchange-correlation functionals. What a more complete determination of energy gaps does require is an exchange-correlation potential which is orbital dependent. 
The polarization and relaxation cloud around an extra electron placed into the conduction bands or a hole created in the valence bands is of different nature than that of the electrons in the ground state of the solid; see, e.\ g., \cite{Fuldebook, FuldeAdv}. The addition or removal of an electron to or from the \textit{infinite} system leaves the electron density unchanged and therefore, the exchange-correlation potential in LDA-DFT, which depends on the local densities only, can not describe properly the correlation hole of the extra charge \cite{FuldeAdv, FuldeInt}. An accurate description of the correlation hole of the extra particle is crucial for determining the band gap in crystals and sets thus the limitations of LDA-DFT.    
Naturally, various improvements have been suggested such as the optimized effective potential method (see, e.g., \cite{Talman, OEP, Stephan}), that allows for the use of orbital-dependent functionals in DFT. 
Orbital-dependent functionals were initially introduced within the exact exchange KS approach \cite{EE}. 
Alternatively, use can  be made of Green's functions techniques, among which the most exploited is the \textit{GW} approximation \cite{Hedin}. Yet, the most enduring problem with the KS approach is that no systematic way has been developed to improve the functionals. 

Presently, advanced quantum chemical wave-function-based methods can address the electron correlation problem at levels of high accuracy in small and medium-size molecular systems. These calculations work usually with virtual excitations of electrons out of molecular orbitals. 
Extending the application of wave-function-based methods to solids is not straightforward since the delocalized nature of the Bloch orbitals would require the consideration of an infinite number of orbitals in order to incorporate the relevant electron correlations. Needless to say, an \textit{ab initio} calculation of the many-electron wave function based on Bloch orbitals scales quite unfavorably with the size of the system. For infinite systems we are concerned with, a relevant issue is to properly account for the translational symmetry while constructing the crystal many-electron wave function \cite{Calais, Resta1, Resta2, Resta3}. 

Despite the associated complexity, wave-function-based algorithms have been developed by several groups and applied to a series of semiconducting and insulating solid-state compounds. The electron correlation problem is addressed in these schemes either by second-order M\o ller-Plesset theory (MP2) \cite{Sun, Ayala, Pisani, Ladik, Hirata, Malrieu} or coupled-electron-pair and coupled cluster methods \cite{Forner, Karin, Hirata2}.

By realizing that the correlation hole of an added electron or hole is a local object, use can be made of local operators \cite{Fuldebook, FuldeInt}. 
The latter are most naturally associated with a set of real-space, localized Wannier orbitals (WO's).    
Localized orbitals are usually obtained by applying various localization schemes to the canonical Bloch orbitals of the self-consistent-field (SCF) ground-state wave function \cite{Pisani, Zikovich, disent, Marzari}. Most often, such algorithms are based on orbital localization procedures 
\cite{Edmiston, Foster-Boys, Pulaypaos} which were developed in order to facilitate correlation treatments in finite systems. 
The Foster-Boys localization criterion \cite{Foster-Boys} was successfully integrated in localization schemes for the canonical Bloch orbitals of periodic systems in several groups, see, e.g., Refs.~\cite{Pisani, Zikovich, disent, Marzari}.

Localized orbital sets are extensively exploited in local correlation methods for solids \cite{Pisani, Stoll1, Gr2, Shukla2, Birken2, Horsch1, Gr1, Albrecht, Birken1, Hozoi, Shukla1, Abdurahman}. A local approach to the computation of inter-atomic and intra-atomic correlations has been suggested by Stollhoff and Fulde \cite{Stollhoff}. Their formalism is based on a physically motivated selection of the relevant two-particle excitations by subdividing the orbital space into appropriate local orbital domains. 
This idea has been elaborated later on by Pulay and Saeb$\o$, 
who developed correlated methods such as local M\o ller-Plesset perturbation theory of second and higher (up to fourth) order without triple excitations and local coupled-electron-pair schemes \cite{Pulay1, Pulay2, Pulay3}. 
These local correlation approaches make use of a \textit{a priori} division of the virtual orbital space into relevant subspaces, i.e., excitation domains. Although developed initially for molecular systems, the local orbital schemes of Pulay and Saeb$\o$ have been also adopted for crystalline compounds; see e.g.\ \cite{Birken2}.
Most recently, the local MP2 correlation approach of Pisani \textit{et al.} \cite{Pisani} has been proposed, which employs linear scaling techniques in order to facilitate the correlation calculation. This formalism was applied to the ground-state properties of various tetrahedral semiconductors. 

A local correlation method which can handle successfully excited-state properties of solids and goes beyond the perturbational schemes is 
the local Hamiltonian approach (LHA) \cite{Gr2, Birken2, Horsch1, Gr1, Albrecht, Birken1, Hozoi}. 
The matrix elements of the local Hamiltonian for the electron-addition and electron-removal quasiparticle states are computed in this approach by means of coupled-cluster or  multiconfiguration and multireference configuration interaction calculations \cite{Helgaker} when the correlations are strong. 
The quasiparticle energy bands are then expressed in terms of those matrix elements \cite{Horsch1, Gr1, Gr2, Birken2, Albrecht, Hozoi, Birken1}. 
The computational effort in the correlation calculation can be drastically reduced by using the method of increments \cite{Stoll1, Stoll2, Paulus}, designed initially to express the ground-state correlation energy as a fast convergent series of correlation contributions assigned to selected groups of localized orbitals. 
The incremental scheme has also become a valuable tool in constructing the scattering operator which is related to the ground-state wave function in a cumulant formulation of the many-body problem \cite{Fuldebook, FuldeAdv, FuldeInt}. 
The LHA has been applied to the determination of the correlation-induced corrections to the band structures of covalent semiconductors and insulators
 \cite{Birken2}, polymers \cite{Birken1}, model beryllium and hydrogen chains \cite{Pahl}, and ionic insulators \cite{Hozoi}. Extensive use of the incremental scheme was made for the accurate determination of the correlation contributions to the excited state properties of diamond and silicon \cite{Gr1, Gr2, Albrecht}.  

The starting point in our LHA is a ground-state periodic HF calculation. It provides a set of canonical Bloch orbitals which are then subject to an orbital localization procedure that yields optimally localized Wannier orbitals; see, e.g., \cite{Zikovich, disent, Forner, Karin}. Alternative approaches were introduced by Shukla \textit{et al.} \cite{Shukla1, Shukla2} and Malrieu \textit{et al.} \cite{Malrieu}, who carried out the ground-state HF calculation in real rather than reciprocal space and obtained directly localized orbitals. 

 \textit{Ab initio} many-body Green's function methods have also been devised to compute the energy bands of periodic systems \cite{Hirata3, Pino, GF1, GF2, GF3}. A local description of the correlation hole around an added electron or hole is utilized in some of these formalisms too \cite{GF1, GF2, GF3, GF4}.
One of the key advantages of such schemes is the fact that the self-energy operator is first evaluated exploiting local orbital sets and then transformed to a momentum representation. In addition, the use of local orbitals prompts to employing the method of increments for the evaluation of the self-energy operator; see, e.\ g., \cite{Fulde, GF4}. 

The key advantage of the wave-function-based methods is the use of well defined and controllable approximations; see, e.g., Ref. ~\onlinecite{Helgaker}. They are amenable to systematic improvements that guarantee converged results for quantities such as binding energies and energy bands \cite{FuldeAdv}. 
 
In the present study, we extend the application of the wave-function-based local Hamiltonian approach to cubic zinc-blende BN (c-BN). 
Due to its peculiar mechanical and physical properties, i.e., high bulk modulus, high thermal conductivity, and low dielectric constant, c-BN is a material of considerable technological interest \cite{Cai}. In addition to experimental investigations of its electronic structure \cite{Fomichev, Chrenko, Agui}, a series of theoretical studies are also available, based either on LDA-DFT \cite{Renata, Park, Rodrigues, Douri, Surh} or HF calculations \cite{Lichanot, Orlando}. 
BN is a good candidate for extending previous studies on ionic oxides \cite{Hozoi} and homonuclear covalent compounds \cite{Gr1, Gr2, Birken2, Albrecht} to the intermediate case of a partially ionic, heteropolar semiconductor.

\section{Theory}
\subsection{Formalism}
We begin with a brief review of the local Hamiltonian formalism and quasiparticle approximation.  
The starting point in our correlation treatment is a periodic HF calculation that provides the self-consistent-field ground-state wave function $|\Phi_{\textsc{scf}}\rangle$ and energy of the \textit{N}-electron system. Here, $|\Phi_{\textsc{scf}}\rangle$ is a single determinant wave function expressed in terms of the Hartree-Fock Bloch orbitals of the infinite system. 
The Bloch functions are eigenstates of the SCF part of the Hamiltonian, $\hat{H}=\hat{H}_{\textsc{scf}}+\hat{H}_{res}$, which is expanded in a Gaussian-type orbital (GTO) basis, $\{f_{i}(\mathbf{r}) \}$, by using proper creation and annihilation operators $a^{\dagger}_{i\sigma}$, $a_{i\sigma}$
\cite{Fulde, FuldeAdv, FuldeInt}:
\begin{eqnarray}
H&=&\sum_{i, j, \sigma}h_{ij}a^{\dagger}_{i\sigma}a_{j\sigma} +
\frac{1}{2}\sum_{\substack {i,j,k,l \\  \sigma, \sigma^{'}}}g_{ijkl}a^{\dagger}_{i\sigma}a^{\dagger}_{k\sigma'}
a_{l\sigma'}a_{j\sigma}
\end{eqnarray}
Here, $h_{ij}$ and $g_{ijkl}$ are one-electron and two-electron matrix elements, respectively. Following the notations in \cite{Fulde}, the compact index \textit{i} of the set of operators $\{ a^{\dagger}_{i\sigma}\}$ consists of an unit cell index \textit{I} and an intracell index \textit{n}. The partition of $\hat{H}$ into a SCF part, $\hat{H}_{\textsc{scf}}$, and a residual interaction part, $\hat{H}_{res}$, is essential within the framework of the local Hamiltonian approach and facilitates the expression of the quasiparticle band energy as a sum of a SCF part and a correlation contribution. 

Next, we consider the HF electron-removal and electron-addition states, described in reciprocal $\textbf{k}$ space as
 \begin{eqnarray}
  \nonumber 
 | \Phi^{N+1}_{\mathbf{k}\nu\sigma}\rangle &=& c_{\mathbf{k}\nu\sigma}^{\dagger} |\Phi_{\textsc{scf}}\rangle  \ \ \  { \mathrm {and}}
 \\ 
 | \Phi^{N-1}_{\mathbf{k'}\mu\sigma'}\rangle &=& c_{\mathbf{k'}\mu\sigma'} |\Phi_{\textsc{scf}}\rangle, 
 \label{scfstates}
 \end{eqnarray}
where $c_{\mathbf{k}\nu\sigma}^{\dagger}$ creates an electron in the conduction-band Bloch state of momentum $\mathbf{k}$, 
band index $\nu$, and spin $\sigma$, and $c_{\mathbf{k}\mu\sigma}$ annihilates an electron in the valence-band Bloch state of momentum $\mathbf{k'}$, band index $\mu$ and spin $\sigma'$.  
The essential idea of the LHA is to derive the correlation corrections to the HF band structure in terms of local matrix elements between explicitly constructed ($N\!+\!1\!$) or ($N\!-\!1\!$) many-electron states. Due to the local character of the correlation hole, these Hamiltonian matrix elements are obtained in practice from calculations on sufficiently large finite clusters.
To accomplish this local description, the one-electron Bloch orbitals are expressed as a superposition of Wannier orbitals associated with the canonical HF valence and conduction bands. The operators $c_{\mathbf{k}\nu\sigma}^{\dagger}$ are expanded in terms of operators for the Wannier orbitals $w^{\dagger}_{\mathbf{R_{I}}n\sigma}$ as 
\begin{eqnarray}
c_{\mathbf{k}\nu\sigma}^{\dagger}=\frac{1}{\sqrt{N_{0}}}\sum_{n, \mathbf{R_{I}}}\alpha_{\nu n}(\mathbf{k})w^{\dagger}_{\mathbf{R_{I}}n\sigma}e^{i\mathbf{k}.\mathbf{R_{I}}}, 
\label{bloch}
\end{eqnarray}
where $\mathbf{R_{I}}$ is the lattice vector of the unit cell \textit{I} at which the Wannier function $|w_{n\sigma} (\mathbf{R_{I}})\rangle$ is centered. $N_{0}$ is the number of unit cells. An analogous expression also applies for $c_{\mathbf{k'}\mu\sigma'}$. Hence, $|\Phi^{N+1}_{\mathbf{k}\nu\sigma}\rangle$ becomes
\begin{eqnarray}
|\Phi^{N+1}_{\mathbf{k}\nu\sigma}\rangle=\frac{1}{\sqrt{N_{0}}}\sum_{n, \mathbf{R_{I}}}\alpha_{\nu n}(\mathbf{k})e^{i\mathbf{k.R_{I}}} w^{\dagger}_{\mathbf{R_{I}}n\sigma} |\Phi_{\textsc{scf}}\rangle
\label{WT}
\end{eqnarray}
and a similar expression holds for $ | \Phi^{N-1}_{\mathbf{k'}\mu\sigma'}\rangle$.
To this end, $|\Phi^{N+1}_{\mathbf{k}\nu\sigma}\rangle$ and $|\Phi^{N-1}_{\mathbf{k'}\mu\sigma'}\rangle$ are subject to a transformation, analogous to the Wannier transformation, which yields local electron-addition and electron-removal one-particle configurations
\begin{eqnarray}
|\Phi^{N+1}_{\mathbf{R_{I}}n\sigma}\rangle&=&w^{\dagger}_{\mathbf{R_{I}}n\sigma} |\Phi_{\textsc{scf}}\rangle,
\nonumber \\
|\Phi^{N-1}_{\mathbf{R_{J}}m\sigma'}\rangle&=&w_{\mathbf{R_{J}}m\sigma'} |\Phi_{\textsc{scf}}\rangle.
\label{rsf} 
\end{eqnarray}
The energies of the conduction bands are expressed in terms of Hamiltonian matrix elements between those local configurations $|\Phi^{N+1}_{\mathbf{R_{I}}n\sigma}\rangle$ \cite{Birken2, Fulde, Fuldebook, FuldeAdv}:
\begin{eqnarray}
\epsilon^{\textsc{scf}}_{\mathbf{k}\nu\sigma}&=&\sum_{\mathbf{R_{I}}}\sum_{nn'}\alpha_{\nu n}(\mathbf{k})\alpha^{*}_{\nu n'}(\mathbf{k}) e^{i\mathbf{k.R_{I}}} H^{\textsc{scf}}_{\mathbf{R_{I}},nn'}, 
\label{scfbands}
\end{eqnarray}
where
\begin{eqnarray}
\nonumber 
H_{\mathbf{R_{I}}-\mathbf{R_{K}}, nn'}^{\textsc{scf}}&=& \langle\Phi_{\textsc{scf}}| w_{\mathbf{R_{K}}n'\sigma}H w^{\dagger}_{\mathbf{R_{I}}n\sigma}|\Phi_{\textsc{scf}}\rangle \\
&-& \delta_{\mathbf{R_{I}}\mathbf{R_{K}}}\delta_{nn'}E^{\textsc{scf}}_{0}.
\label{LHSCF}
\end{eqnarray}
These matrix elements are uniquely defined. Here, $E_{0}^{\textsc{scf}}$ is the Hartree-Fock ground-state energy. An expression analogous to Eq. (\ref{scfbands}) applies for the energies of the valence bands.
For the valence-band states, the matrix elements have the form
 \begin{eqnarray}
\nonumber
H_{\mathbf{R_{J}}-\mathbf{R_{L}}, mm'}^{\textsc{scf}}&=& \delta_{\mathbf{R_{J}}\mathbf{R_{L}}}\delta_{mm'}E^{\textsc{scf}}_{0} \\
&-&\langle\Phi_{\textsc{scf}}| w^{\dag}_{\mathbf{R_{L}}m'\sigma'}H w_{\mathbf{R_{J}}m\sigma'}|\Phi_{\textsc{scf}}\rangle.
\label{LHSCF2}
\end{eqnarray}
The local Hamiltonian matrix elements in Eqs. (\ref{LHSCF}) and (\ref{LHSCF2}) are extracted from calculations on large enough clusters \cite{Birken1, Birken2, Hozoi}. 

Next, we consider the inclusion of electron correlations in the band structure calculations. For relatively weakly correlated electron systems, such as boron nitride, we may adopt the quasiparticle approximation \cite{Fulde, Fuldebook, Birken2}. Within this approximation, the life time of the excitations is neglected. Satellite structures are also beyond the quasiparticle picture. 
In analogy with the HF approximation, the correlated conduction and valence energy bands are expressed as 
\begin{eqnarray}
\nonumber
\epsilon_{\mathbf{k}\nu\sigma}&=& \langle \Psi_{\mathbf{k}\nu\sigma}^{N+1}|H|\Psi_{\mathbf{k}\nu\sigma}^{N+1}\rangle-E_{0} \ \ \  { \mathrm {and}}
\\
\epsilon_{\mathbf{k'}\mu\sigma'}&=&E_{0}-\langle \Psi_{\mathbf{k'}\mu\sigma'}^{N-1}|H|\Psi_{\mathbf{k'}\mu\sigma'}^{N-1}\rangle, 
\end{eqnarray}
where $E_{0}$ is the energy of the correlated ground state and $| \Psi_{\mathbf{k}\nu\sigma}^{N+1}\rangle$ and $|\Psi_{\mathbf{k'}\mu\sigma'}^{N-1}\rangle$ are the correlated counterparts of $|\Phi_{\mathbf{k}\nu\sigma}^{N+1}\rangle$ and $| \Phi_{\mathbf{k'}\mu\sigma'}^{N-1}\rangle$, respectively.  
If a Wannier-like transformation is applied to the wave functions $|\Psi_{\mathbf{k}\nu\sigma}^{N+1}\rangle$ and $|\Psi_{\mathbf{k'}\mu\sigma'}^{N-1}\rangle$, the locally correlated wave functions $|\Psi_{\mathbf{R_{I}}n\sigma}^{N+1}\rangle$ and $|\Psi_{\mathbf{R_{J}}m\sigma'}^{N-1}\rangle$ are obtained. 
The latter are viewed as derived from the correlated ground-state wave function, after an electron-addition to the conduction band or an electron-removal process from the valence band has taken place and the subsequent response of the crystalline surroundings to the additional charge is explicitly accounted for \cite{Gr1, Gr2, Albrecht, Fuldebook, Birken2, Hozoi}.  

In analogy with the \textsc{HF} case, the energy bands are expressed in terms of real-space matrix elements between localized ($N\!\pm\!1\!$) states. In this case, however, the matrix elements incorporate electron correlation effects.
The conduction energy bands are given by
\begin{eqnarray} 
\epsilon_{\mathbf{k}\nu\sigma}& =&\sum_{\mathbf{R_{I}}}\sum_{nn'}\alpha_{\nu n}(\mathbf{k})\alpha^{*}_{\nu n'}(\mathbf{k}) e^{i\mathbf{k}.\mathbf{R_I}} 
\nonumber \\
&\times&
\langle\Psi_{\mathbf{0}n'\sigma}^{N+1}|H|\Psi_{\mathbf{R_{I}}n\sigma}^{N+1}\rangle-E_{0} \delta_{\mathbf{0R_{I}}}\delta_{nn'}
\nonumber \\
&=&\epsilon^{\textsc{scf}}_{\mathbf{k}\nu\sigma}+\epsilon^{corr}_{\mathbf{k}\nu\sigma}.
\label{corbands}
\end{eqnarray}
The expression for the correlated valence bands is analogous.  
The expressions for the energy bands can also be written in terms of cumulants, which insures size-extensivity of the different correlation energy contributions. The cumulant formulation has been discussed in detail in Refs.  ~\onlinecite{FuldeAdv} and ~\onlinecite{Fuldebook}.   

In real space, we can represent the \textit{correlated} ($N\!+\!1\!$) electron states as
\begin{eqnarray}
|\Psi_{\mathbf{R_{I}}n\sigma}^{N+1}\rangle=e^{S}|\Phi_{\mathbf{R_{I}}n\sigma}^{N+1} \rangle,
\end{eqnarray}
and similarly for the localized and correlated ($N\!-\!1\!$) electron states. The operator \textit{S} is constructed from a selected set of excitation operators; see, e.\! g., \cite{Fuldebook, Kladko, Gr1, Gr2}. 
For relatively weakly correlated systems, a reasonable ansatz for the operator \textit{S} includes one- and two-particle excitation operators \cite{Gr1, Gr2, Fuldebook, Birken2},  
\begin{eqnarray} 
\nonumber 
S&=&\sum_{\mathbf{R_J, R_K}}\sum_{a,m}\sum_{\sigma,\sigma'} \eta^{am}_{\mathbf{R_JR_K}} a^{\dag}_{\mathbf{R_{J}}a\sigma'} w_{\mathbf{R_{K}}m\sigma'}w^{\dag}_{\mathbf{R_{I}}n\sigma} w_{\mathbf{R_{I}}n\sigma} \\
\nonumber 
& +&\sum_{\substack{\mathbf{R_J, R_K} \\ \mathbf{R_L, R_M}} }\sum_{\substack{a,b \\ m, m'}}\sum_{\sigma,\sigma'}\eta^{abmm'}_{\mathbf{R_JR_KR_LR_M}} \\
&\times&a^{\dag}_{\mathbf{R_{J}}a\sigma}a^{\dag}_{\mathbf{R_{K}}b\sigma'}w_{\mathbf{R_{L}}m\sigma'} w_{\mathbf{R_{M}}m'\sigma},
\label{S}
\end{eqnarray}
where the indices \textit{m, m'} refer to the occupied Wannier orbitals of the ($N\!+\!1\!$) electron system and $a, b$ denote virtual orbitals  centered at $\mathbf{R_J}$,  $\mathbf{R_K}$. The expression for the operator \textit{S} for the ($N\!-\!1\!$) electron system is analogous. 

The operator $\textit{S}$ has a twofold function. The first term in Eq. (\ref{S}) generates the relaxation and polarization cloud around an extra electron, created by the composite operator $w^{\dag}_{\mathbf{R_{I}}n\sigma} w_{\mathbf{R_{I}}n\sigma}c^{\dag}_{\mathbf{k}\nu\sigma}$ in the conduction-band Wannier orbital $|w_{n\sigma}(\mathbf{R_{I}})\rangle$. Analogously, a relaxation and polarization cloud is formed around a hole created by $w_{\mathbf{R_{J}}m\sigma'}w^{\dag}_{\mathbf{R_{J}}m\sigma'}c_{\mathbf{k'}\mu\sigma'}$ in the valence-band WO $|w_{m\sigma'}(\mathbf{R_{J}})\rangle$. The second term in Eq. (\ref{S}) accounts for the so-called loss of ground- state correlation \cite{Fulde, Fuldebook}. We describe first the procedure for the computation of the first term. 

On-site and short-range inter-atomic relaxation and polarization effects are computed by means of separate SCF orbital optimizations for a large but finite region around the additional charge. The frozen local hole approximation \cite{Birken2, Pahl} is applied, i.e., the Wannier orbital where the extra hole or electron resides is kept frozen in the SCF calculation \cite{MCSCF} for the ($N\!\pm\!1\!$) excited state. 
This approximation is well suited if the electrons are moderately correlated. 
It has been demonstrated \cite{Pahl} that up to first-order perturbation theory, the SCF orbital relaxations are equivalent to the one-particle excitations around the frozen Wannier orbital, see the first term in Eq. (\ref{S}). The corresponding correlated wave functions $|\tilde{\Psi}_{\mathbf{R_{J}}m\sigma'}^{N-1}\rangle$ and $|\tilde{\Psi}_{\mathbf{R_{I}}n\sigma}^{N+1}\rangle$ \footnote{The wave functions $|\tilde{\Psi}_{\mathbf{R_{J}}m\sigma'}^{N-1}\rangle$ and $|\tilde{\Psi}_{\mathbf{R_{I}}n\sigma}^{N+1}\rangle$ are correlated only through one-particle excitations.} are thus identical up to first order to the optimized SCF states $|\tilde{\Phi}_{\mathbf{R_{J}}m\sigma'}^{N-1}\rangle$ and $|\tilde{\Phi}_{\mathbf{R_{I}}n\sigma}^{N+1}\rangle$, respectively.    
Note, that the creation and annihilation operators in Eq. (\ref{S}) refer to \textit{spin-orbitals}. In practice, a \textit{spatial orbital} set is exploited and therefore we also allow for the relaxation of the orbital accommodating the extra electron or hole within the computed relaxed surrounding in order to account for the spin degree of freedom; see Section IV. 
Wave functions which are expressed in terms of relaxed orbitals within the nearby surroundings and readjusted hole or added-electron orbitals are denoted in the following as $|\breve{\Phi}_{\mathbf{R_{J}}m\sigma'}^{N-1}\rangle$ and $|\breve{\Phi}_{\mathbf{R_{I}}n\sigma}^{N+1}\rangle$.    

The long-range polarization of the crystal brings also large corrections to the diagonal matrix elements of the local Hamiltonian and is estimated within the approximation of a dielectric continuum \cite{Gr1, Fuldebook, Birken2, Hozoi}. 

Next, we consider the loss of ground-state correlation. 
This effect is related to correlation contributions that are absent in the ($N\pm1$) electron system but present in the $N$-electron ground state. This has to do with the fact that some excitations involving the removed electron or the orbital where the extra electron is placed are blocked in the ($N\pm1$) electron system.

To account for electron correlation effects beyond relaxation and polarization, an important part of which is the loss of ground-state correlations, multi-reference (MR) single and double configuration interaction (SDCI) calculations \cite{ICCI, ICCI2} are carried out for each $|\tilde{\Phi}_{\mathbf{R_{J}}m\sigma'}^{N-1}\rangle$ ($|\breve{\Phi}_{\mathbf{R_{J}}m\sigma'}^{N-1}\rangle$) and $|\tilde{\Phi}_{\mathbf{R_{I}}n\sigma}^{N+1}\rangle$ ($|\breve{\Phi}_{\mathbf{R_{I}}n\sigma}^{N+1}\rangle$) state. 
The CI wave functions, $|\Psi_{\mathbf{R_{J}}m\sigma'}^{N-1}\rangle$ and $|\Psi_{\mathbf{R_{I}}n\sigma}^{N+1}\rangle$, are constructed within an orbital space consisting of the relaxed doubly occupied SCF orbitals, the singly occupied, $|w_{m\sigma'}(\mathbf{R_{J}})\rangle$ or $|w_{n\sigma}(\mathbf{R_{I}})\rangle$, and the set of virtual orbitals. 
Within the quasiparticle approximation, the CI expansions contain only configurations in which the hole in $|w_{m\sigma'}(\mathbf{R_{J}})\rangle$ or the extra electron in $|w_{n\sigma}(\mathbf{R_{I}})\rangle$ is never promoted to another orbital.
Differential correlation effects arising from having a different number of electrons in the $N$ and ($N\!\pm\!1\!$) configurations are also assessed by means of CI calculations. 

\subsection{Calculation of matrix elements}
The required matrix elements can be obtained from cluster calculations, as pointed out above. The crystalline environment around such a cluster is described by an embedding potential
which is extracted from the periodic Hartree-Fock calculation.
This embedding scheme was proposed by Birkenheuer \textit{et al.}, see Refs.~\onlinecite{Birken2}
and \onlinecite{Embed}, and requires a prior localization of the canonical HF orbitals
of the $N$-electron periodic system.
The practical evaluation of the embedding potential is facilitated by the knowledge of the
 crystal Fock operator $F_{cryst}$ and the cluster Fock operator $F_{clust}[P_{clust}]$.
The latter is related to the density operator 
$P_{clust}=\sum_{m\sigma}^{occ, clust} |\tilde{w}_{m\sigma} \rangle\langle \tilde{w}_{m\sigma}|$,
where $|\tilde{w}_{m\sigma}\rangle$ are the localized occupied orbitals within the cluster.

The cluster is subdivided into an active region and a buffer region.
The occupied and low-lying conduction-band orbitals of the active region are
explicitly correlated in the post-HF calculation, while those of the buffer region are 
always kept frozen.
The role of the atoms in the buffer region is to insure an accurate representation of the
longer-range tails of the Wannier orbitals, $|w_{m\sigma'}(\mathbf{R_{J}})\rangle$ and
$|w_{n\sigma}(\mathbf{R_{I}})\rangle$, centered at the atomic sites J and I of the active region; see \cite{Birken2, Hozoi} and also Section IV.

The orbital set associated with the finite cluster is generated from the original 
crystal Wannier orbitals.
The latter are obtained via a Wannier-Boys localization scheme \cite{Zikovich, Baranek}
 for the core, valence, and low-lying conduction-band states. Further, the crystal WO's are projected onto the set of atomic basis functions assigned to the cluster sites \cite{Birken2, Hozoi}.  
These \textit{projected} Wannier orbitals, $|w'_{m\sigma'}(\mathbf{R_{J}})\rangle$ and
$|w'_{n\sigma}(\mathbf{R_{I}})\rangle$, do not constitute an orthonormal orbital set. 
Therefore, a series of groupwise orthonormalizations are carried out for the core, valence, and low-lying
conduction-band orbital groups, both for the active and buffer regions. The sets of orthonormalized occupied and low-lying conduction-band
orbitals are denoted as $|w''_{m\sigma'}(\mathbf{R_{J}})\rangle$ and $|w''_{n\sigma}(\mathbf{R_{I}})\rangle$. 

Additionally, a set of virtual orbitals is needed in order to construct the complete variational orbital space for the correlation calculation. To generate those virtual orbitals, a modified version of the projected atomic orbital (PAO) formalism of Hampel and Werner \cite{Hampel} is adopted \cite{Birken2, Embed}. The idea for using projected atomic orbitals as a local virtual basis was initially proposed by Pulay \cite{Pulaypaos}.  

The virtual PAO's are obtained from the atomic orbital basis of the active cluster region
after projecting out the contribution of the orthonormalized, occupied and low-lying conduction-band
orbitals $|w''_{m\sigma'}(\mathbf{R_{J}})\rangle$ and $|w''_{n\sigma}(\mathbf{R_{I}})\rangle$ via a Schmidt orthogonalization scheme; see also Ref. \onlinecite{Hozoi}.
The PAO's are subsequently L\"owdin orthonormalized. The final variational orbital space of the embedded cluster consists of the  core, valence-band, and low-lying conduction-band orbitals $|w''_{m\sigma'}(\mathbf{R_{J}})\rangle$ and $|w''_{n\sigma}(\mathbf{R_{I}})\rangle$ together with the set of virtual PAO's.
The local electron-addition and electron-removal configurations,
$|\Phi_{\mathbf{R_{I}}n\sigma}^{N+1}\rangle$ and $|\Phi_{\mathbf{R_{J}}m\sigma'}^{N-1}\rangle$, are defined
within this variational orbital space. 
\section{Crystal Structure of c-BN and Computational information}
Boron nitride belongs to the group of III-V binary compounds and exhibits three solid state phases.
At room temperature and normal atmospheric pressure it crystallizes in a hexagonal structure,
which transforms to denser packed zinc-blende or wurtzite structures under static
\cite{Bundy, Wakatsuki} or dynamic compression\cite{Soma}, respectively. 
The cubic zinc-blende form is face-centered (fcc), with the $F\bar{4}3m$ space 
group symmetry. 
The corresponding lattice constant is 3.615 \AA \, \cite{Wentorf}.
The local point-group symmetry for both B and N is $T_{d}$, with a four site nearest-neighbor
coordination.

GTO basis sets of triple-zeta quality augmented with polarization functions were applied for
the N atoms. 
We used as a starting point the N 7-311G GTO basis set derived by Dovesi
\textit{et al.} \cite{Dovesi} for crystalline LiN.
We added to Dovesi's basis set a single-Gaussian $d$ polarization function with an exponent of 0.8
\cite{Orlando, Dovesi2} and reoptimized the exponents of the outermost two $sp$ shells.
The exponents of the reoptimized outermost $sp$ shells are $0.4523$ and
$0.220241$.
Two different GTO basis sets were used for the B atoms. 
The first is a triple-zeta B 6-311G* basis set which is derived from the B 6-21G* GTO basis set constructed for BN by Orlando \textit{et al.} \cite{Orlando}. This 6-21G* basis set was suitably modified for the purpose of obtaining a triple-zeta basis set. 
The optimized values of the exponents and contraction coefficients of the valence $2sp$ shell and the exponents of the polarization single-Gaussian $3sp$ and $4sp$ shells are listed in
Table \ref{BS1} of the Appendix.

The second basis set is a double-zeta 6-41G  basis set which is designed starting from the 6-311G* basis set given in Table  \ref{BS1}. This smaller basis set is employed for calculations on very large clusters, for which the use of the  6-311G* basis is at present computationally unfeasible. Those large clusters are selected 
in order to compute additional increments. 
As discussed in the following section, the associated basis set effect is either negligible or very small.
To obtain the 6-41G basis set, the \textit{d} polarization function is removed and the single-Gaussian $3sp$ function is contracted with the three primitives of the $2sp$ shell. The exponents and contraction coefficients of the $2sp$ shell and the exponent of the new single-Gaussian $3sp$ function are reoptimized. Their values are provided in Table \ref{BS2} of the Appendix. 

The periodic HF calculations are performed with the \textsc{crystal} program package
\cite{CRYSTAL1}.
\textsc{crystal} provides in addition a Wannier-Boys orbital localization module 
for generating real-space, localized Wannier functions \cite{Zikovich}.
The set of projected Wannier orbitals for the finite cluster and the matrix representation of the
crystal Fock operator $F_{cryst}$ in terms of the cluster atomic basis functions
are both obtained with the \textsc{crystal-molpro} interface program \cite{interface}.
The embedding potential for the finite cluster is generated with \textsc{molpro}
\cite{molpro} by substracting from $F_{cryst}$ the cluster Fock operator 
$F_{clust}[P_{clust}]$, the latter being computed from the density matrix associated
with the set of occupied, projected WO's. 
All subsequent calculations for studying the effects of electron correlations are
performed with the \textsc{molpro} package \cite{molpro}. 
\section{Correlation induced corrections to the band structure of zinc-blende c-BN}
\begin{figure}[htbp]
\begin{center}
\includegraphics[width=8.5cm]{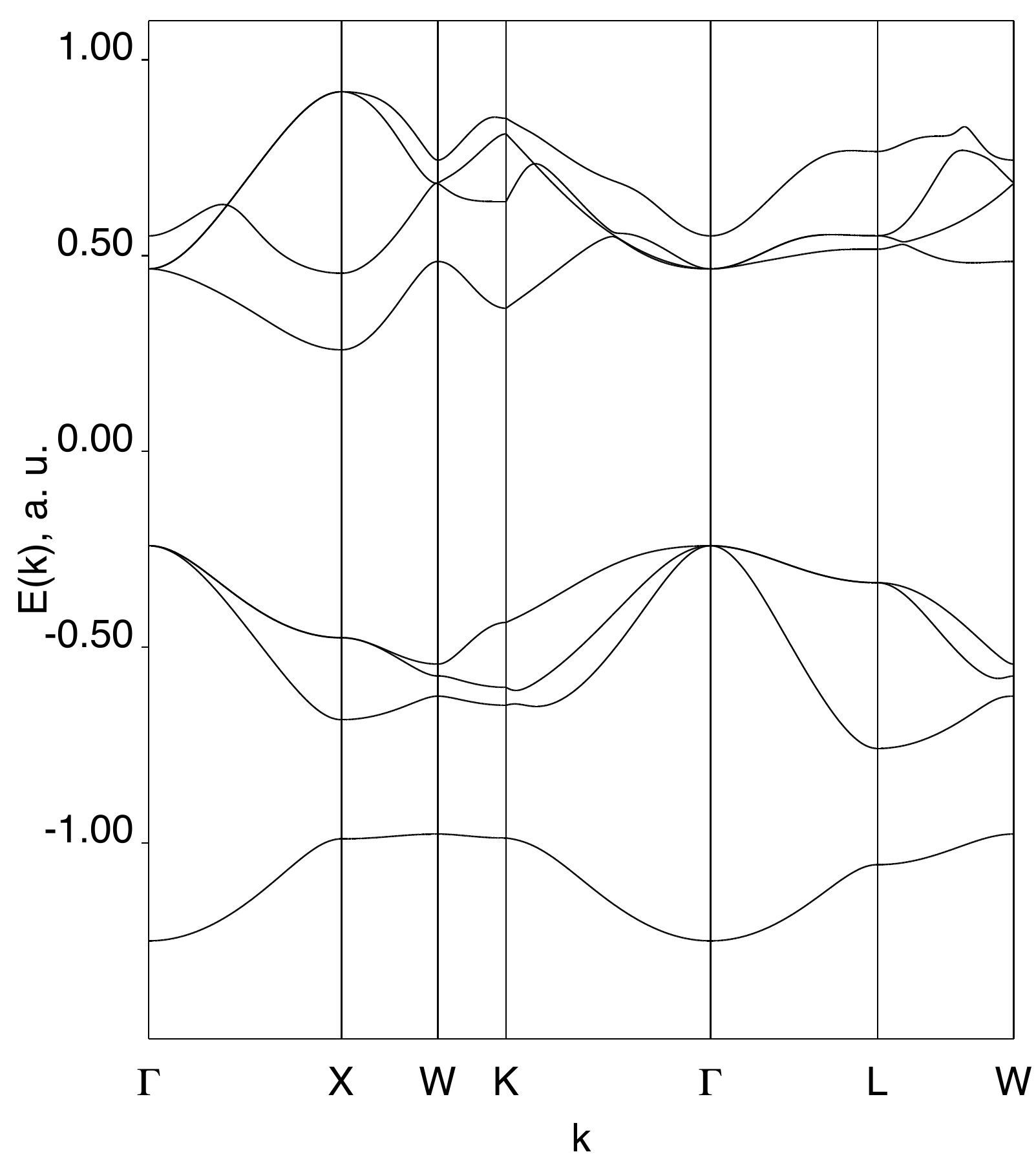}
\caption{Hartree-Fock band structure of c-BN. The core N $1s$ and B $1s$ bands are not shown in the figure.}
\label{hfband}
\end{center}
\end{figure}

The HF energy bands of c-BN, obtained with the triple-zeta basis sets described in 
the previous section, are plotted in Fig. \ref{hfband}.
The HF energy per unit cell is $-79.268$ a.u., close to the
value obtained by Euwema \textit{et al.}, $-79.250$ a.u. \cite{Euwema}.
The upper valence bands of c-BN have N $2p$ character, with small
contributions from the B $2s$ and $2p$ states.
The low-lying conduction bands have predominant B $2s$ and B $2p$ character, with sizeable  
admixture from the N $2s$,$3s$ and $2p$,$3p$ functions.
The fundamental band gap of the system is indirect, $\Gamma^{v}_{15} \rightarrow X^{c}_{1}$.
At the Hartree-Fock level this quantity is strongly overestimated, i.e., 13.62 eV, and therefore 7 eV larger 
than experimental values deduced from soft x-ray experiments at the B and N $K$-edges \cite{Fomichev, Agui} and 
optical absorption \cite{Chrenko} measurements. Those values are 6.0$\pm$0.5 and 6.4$\pm$0.5 eV, respectively.
The energy separation between the N $2p$ and lower-lying N $2s$ valence bands is
about 6 eV.
\begin{figure}[htbp]
\begin{center}
\includegraphics[width=8.5cm]{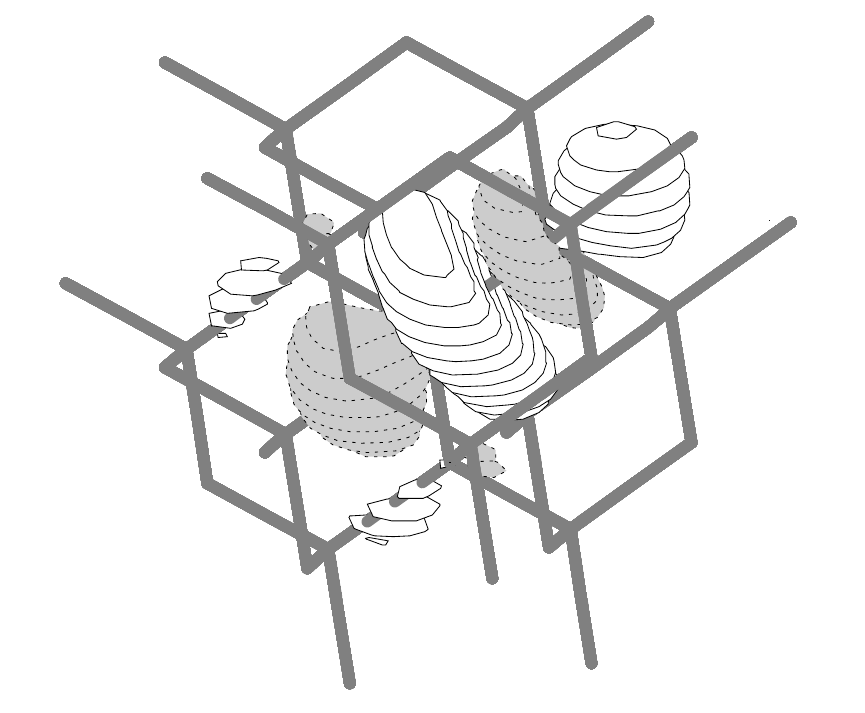}
\caption{One of the four conduction-band $sp^{3}$-like Wannier orbitals with predominant B $2s, 2p$ character after projection onto a [B$_{13}$N$_{28}$] cluster.}
\label{WF0}
\end{center}
\end{figure}

Wannier-Boys transformations are carried out separately for the core, valence, and low-lying
conduction bands.
The lowest four conduction bands of BN are separated by a small gap from the higher
virtual Bloch states.
Due to this finite energy gap, no numerical problems occur in the Wannier-Boys transformation
for the set of low-lying conduction-band states. 
The resulting WO's associated with these lowest conduction bands turn out to be
a set of four $sp^{3}$-like hybrids, each oriented along one of the B-N segments of a
BN$_{4}$ tetrahedron.
They have large weight at the nearest-neighbor N
sites and strong antibonding character with respect to the nearest-neighbor N $2s$ and N $2p$ orbitals. 
Such a Wannier orbital, projected onto a finite cluster, is plotted in Fig.~\ref{WF0}.

\begin{figure}[htbp]
\begin{center}
\includegraphics[width=8.5cm]{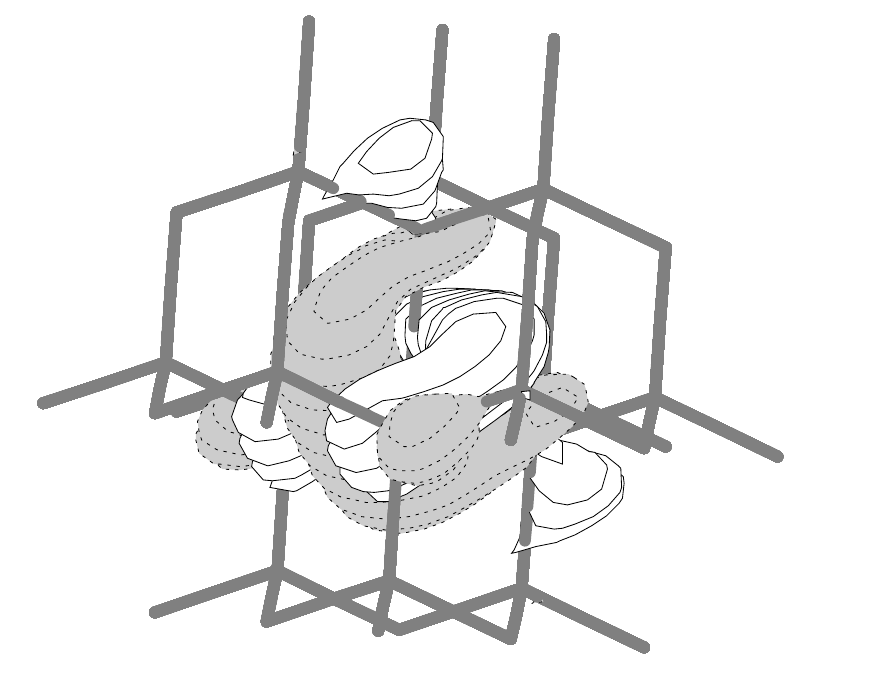}
\caption{B $2p$-like conduction-band Wannier orbital after projection onto a [B$_{13}$N$_{28}$] cluster.}
\label{WF2}
\end{center}
\end{figure}

\begin{figure}[htbp]
\begin{center}
\includegraphics[width=8.5cm]{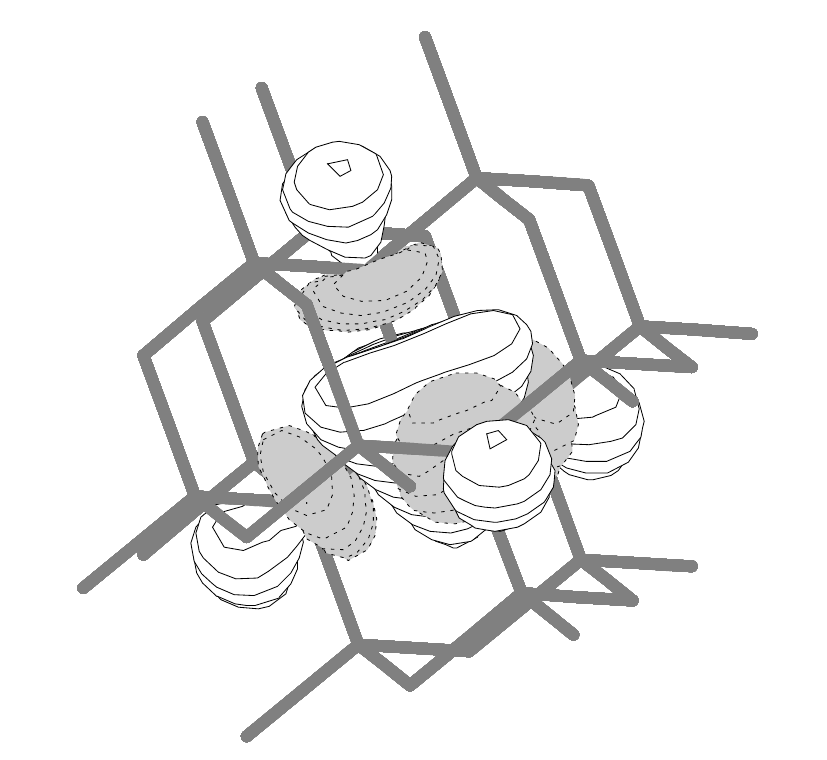}
\caption{B $2s$-like conduction-band Wannier orbital after projection onto a [B$_{13}$N$_{28}$] cluster. }
\label{WF1}
\end{center}
\end{figure}

In a next step, the projected WO's are subject to a Pipek-Mezey 
transformation \cite{PM}.
In simple cases, this procedure transforms a set of $sp$ hybrids into 
Cartesian type, $s$ and $p$ functions, as found for example in the case of the
conduction-band Wannier orbitals of MgO \cite{Hozoi}.  
Here, the three degenerate orbitals, each oriented along one of the $C_2$ axes,
and the fourth, lower-energy WO obtained after the Pipek-Mezey transformation
bear little resemblance to Cartesian functions.
Each $p$-like lobe of the three degenerate orbitals at the B site of a BN$_{4}$ tetrahedron
is antibonding with respect to the two N ions adjacent to that lobe, but has substantial, 
U-shaped bonding tails at the other two N sites \footnote{For each of these $p$-like lobes, two of the nitrogen atoms of a BN$_{4}$ tetrahedron are closer as compared to the other two N sites, see Fig. \ref{WF2}.}.
The fourth function is also strongly deformed as compared to a pure $s$ function. 
In the immediate neighborhood of a B site, it is tetrahedrally shaped.
These projected WO's, obtained through the Pipek-Mezey transformation,
are plotted in Figs. ~\ref{WF2} and ~\ref{WF1}.  

\begin{figure}[htbp]
\begin{center}
\includegraphics[width=8.5cm]{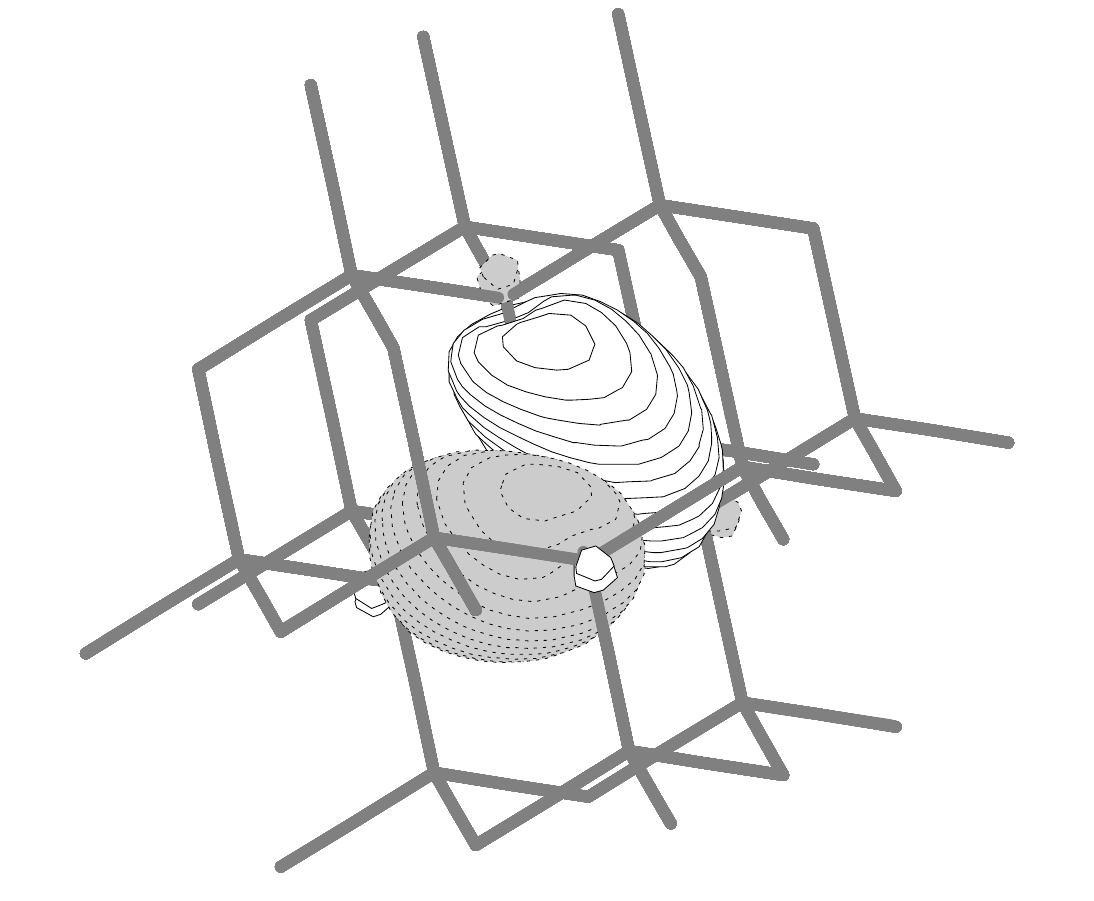}
\caption{N $2p$-like valence-band Wannier orbital, projected onto a [N$_{13}$B$_{28}$] cluster.}
\label{WF3}
\end{center}
\end{figure}

\begin{figure}[htbp]
\begin{center}
\includegraphics[width=8.5cm]{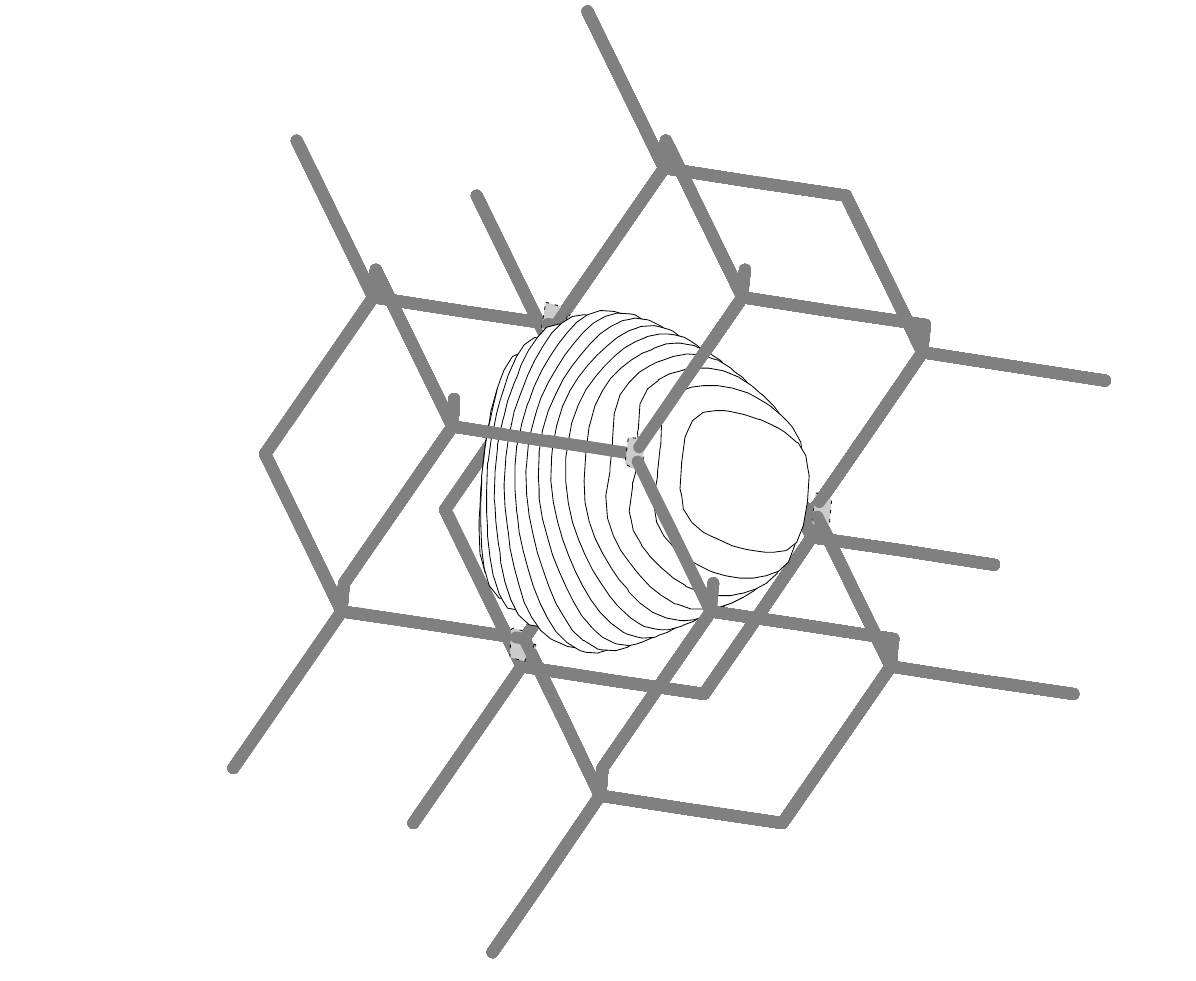}
\caption{N $2s$-like valence-band Wannier orbital, projected onto a [N$_{13}$B$_{28}$] cluster. Due to the local environment, it is distorted to a tetrahedral-like shape.}
\label{WF4}
\end{center}
\end{figure}

In contrast to the conduction-band Wannier functions, the Wannier orbitals associated with the
nitrogen $2s$ and $2p$ valence bands are well localized 
around the N sites, with very small tails at the nearest boron neighbors.
The Pipek-Mezey transformation yields a set of somewhat deformed, Cartesian-like $2s$,
$2p_{x}$, $2p_{y}$, and $2p_{z}$ functions.
One of the $2p$-like components and the deformed N $2s$ orbital are plotted in 
Fig.~\ref{WF3} and Fig.~\ref{WF4}, respectively. 
Worthwhile to mention, the norms of the projected valence-band WO's, centered within the active regions of
the embedded clusters, are always larger than 0.99 of the initial HF WO's obtained
with {\sc crystal}. The corresponding norms for the projected conduction-band WO's are always larger than
0.94. 

\begin{figure}[htbp]
\begin{center}
\includegraphics[width=8.5cm]{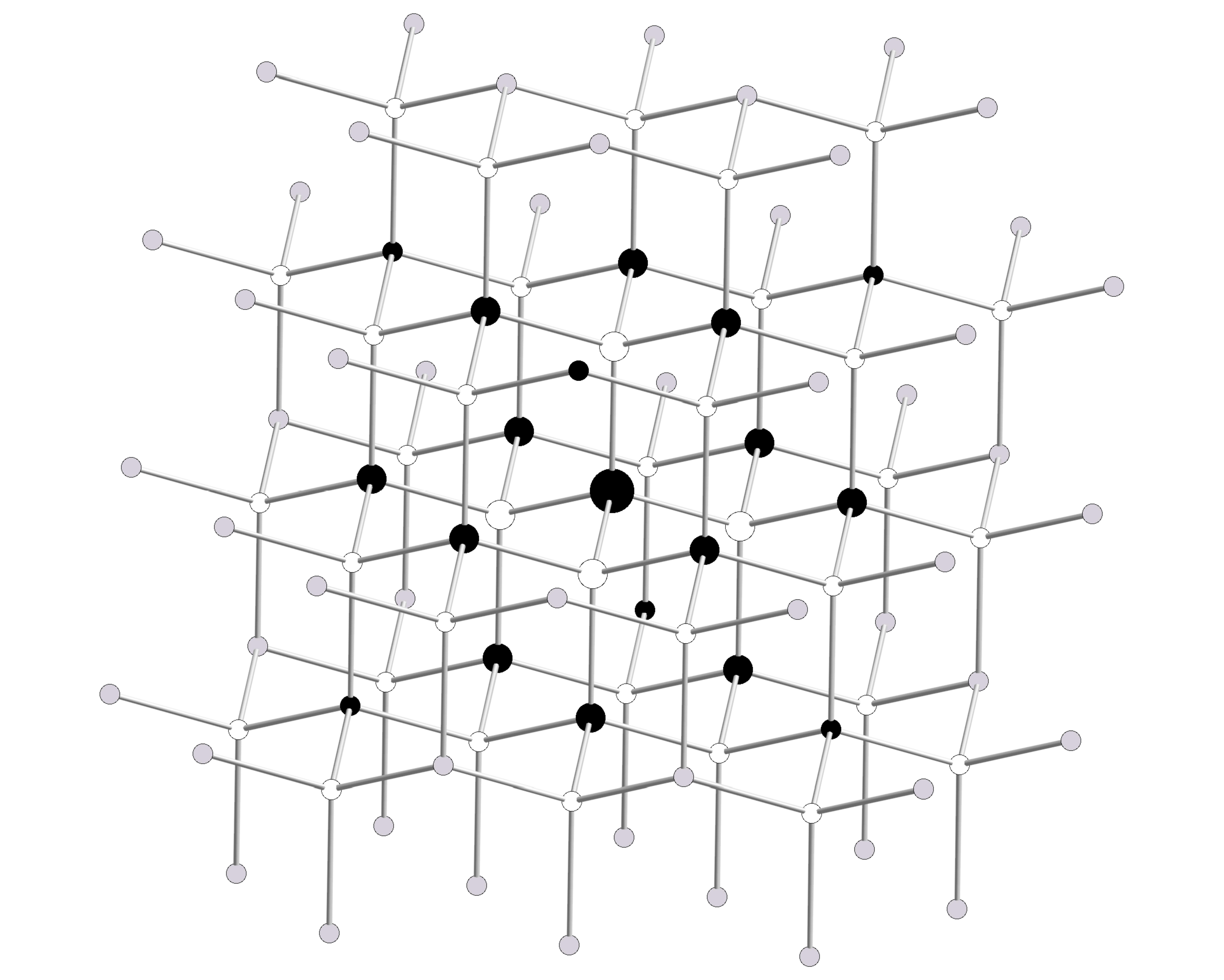}
\caption{Illustration of the [N$_{71}$B$_{36}$] embedded cluster. The active region of the cluster consists of a [N$_{17}$B$_{4}$] fragment. The N sites in that region are depicted as black spheres of different size, depending on their distance from the central N site. The latter is shown as the largest black sphere. N atoms in the buffer zone are shown as small, grey spheres. The four active B atoms are drawn as large white spheres, whereas the buffer B sites are shown as small white spheres.}
\label{cluster}
\end{center}
\end{figure}
\subsection{Diagonal matrix elements of the local Hamiltonian}
\subsubsection{Short-range relaxation and polarization effects}
We start by considering the short-range correlation-induced corrections to the diagonal matrix elements of the Hamiltonian in Eqs. (\ref{LHSCF}) and (\ref{LHSCF2}). We construct first the correlation cloud of a hole introduced in the valence bands.
To evaluate the short-range part of this correlation cloud, i.e., the on-site relaxation as well as the relaxation and polarization effects associated with the first (nearest-neighbor, nn) and second (next-nearest-neighbor, nnn) nitrogen coordination shells around the N site where the hole resides, we designed three different clusters:[N$_{55}$B$_{28}$], [N$_{71}$B$_{36}$], and [N$_{63}$B$_{44}$]. 
The first nitrogen coordination shell contains 12 sites whereas the second shell consists of 6 N atoms. 

The [N$_{55}$B$_{28}$] cluster is employed for computing the on-site and nn relaxation and polarization effects.  
The active region, [N$_{13}$B$_{4}$], of the [N$_{55}$B$_{28}$] cluster includes a central N site, sketched as the largest black sphere in Fig.\  \ref{cluster}, the four nn B atoms, depicted as large white spheres, and the first shell of 12 nn N atoms. 
These 12 N nn's are drawn as average-size black spheres in Fig. \ref{cluster}. The buffer region around the active subunit contains all B and N neighbors in the first two coordination shells of the 12 outer N sites of the active region. Such a buffer region is large enough to ensure a good representation of the long-range tails of the Wannier orbitals centered in the active subunit. 
In this set of calculations, we exploit the 6-41G basis set for the B atoms (see Section III) because the use of the 6-311G* basis set is computationally unfeasible for such a large cluster. The nitrogens are described by the 7-311G* basis set.  

Separate restricted open-shell HF (ROHF) calculations are carried out for the N $2s$ and N $2p$ ($N\!-\!1\!$) electron states. First, the local electron-removal states $|\Phi^{N-1}_{\mathbf{R_{J}}m\sigma'}\rangle$ are constructed according to Koopmans' theorem. 
In a second step, the hole is placed in one of the N $2s$ or $2p$ WO's at the central N site of the [N$_{55}$B$_{28}$] cluster and the other valence orbitals within the active region are allowed to relax and polarize in response to
the electron-removal process. 
The Wannier orbital with the hole is kept frozen\cite{Birken2, Hozoi}. We construct thus $|\tilde{\Phi}^{N-1}_{\mathbf{R_{J}}m\sigma'}\rangle$ as a first approximation to the correlated wave function $|\Psi^{N-1}_{\mathbf{R_{J}}m\sigma'}\rangle$; see also Section II.

The corrections due to on-site orbital relaxation are found to be substantial, 1.3--1.6 eV for the N $2s$ and $2p$ hole states (first line in Table \ref{diagonalhole}). The effect is stronger for the N $2s$ hole state because in this case the on-site relaxation involves all six electrons in the highly polarizable N $2p$ shell. 
The relaxation of the nn boron $1s$ core shells is only a small fraction of 0.01 eV of the overall correlation-induced shift of the valence bands, as shown on the second line in Table \ref{diagonalhole}. Therefore this effect is not discussed further.
The relaxation and polarization of the nn N orbitals bring corrections of similar magnitude for both types of hole states, about 1.6 eV (third entry in Table \ref{diagonalhole}).
 The relaxation of the nitrogen $1s$ core shell is vanishingly small. For this reason, all N core shells are kept frozen throughout the calculations.

  \begin {table}
 \centering 
 \caption{Correlation-induced corrections (in eV) to the diagonal Hamiltonian matrix elements for the valence-band N $2s$ and $2p$ hole states. Negative corrections indicate an upward shift of the valence bands.}
\begin{tabular}{p{5cm}p{1.6cm}p{1.6cm}}
  \hline\hline
 &$\Delta$ H$_{mm}$($\textbf{0}$)& \\ \hline 
&N $2p$&N $2s$\\ \hline
On-site orb. relaxation&-1.29&-1.64\\
nn(B) orb. relaxation &-0.01&-0.01\\
nn(N) orb. relaxation &-1.57& -1.63\\
nnn(N) orb. relaxation & -0.14 & -0.14 \\
Hole orb. relaxation &-0.18&-0.12 \\
Loss of ground-state corr. &1.05&1.08 \\
Long-range polarization &-1.53& -1.53\\
Total corr. correction &\textbf{-3.67}&-3.99 \\ 
 \hline \hline
 \end{tabular} 
\label{diagonalhole}
\end {table}

Some comments regarding the basis set used for the B atoms are here in place. Basis set effects were investigated by additional calculations for the on-site relaxation and polarization using a [N$_{13}$B$_{28}$] cluster.
The active region, [NB$_{4}$], of the [N$_{13}$B$_{28}$] cluster contains only the central N site and the four nn B atoms. The first shell of twelve nn nitrogens around the central N site is now part of the buffer region.  
ROHF calculations are performed for the N $2s$ and N $2p$ ($N\!-\!1\!$) electron states of the [N$_{13}$B$_{28}$] cluster using for the borons either 6-41G or 6-311G* basis sets. 
The basis set effect for the on-site orbital relaxation is found to be negligibly small, $<$1\%, for both the N $2s$ and  and $2p$ hole states. 

In a next step, the contribution of the nnn nitrogens to the overall relaxation and polarization has been investigated. Since wave-function-based calculations for a cluster containing all nn and nnn N sites are computationally unfeasible even with the double-zeta basis set, we designed clusters where only part of the six nnn N sites were included in the active region. 
For one of these clusters, [N$_{71}$B$_{36}$], the active region [N$_{17}$B$_{4}$] includes the active region [N$_{13}$B$_{4}$] of the [N$_{55}$B$_{28}$] cluster and four nitrogens from the second coordination shell of the central N site. These four nnn N's lie along the $C_{2y}$ and $C_{2z}$ axes.
Analogously, the active region [N$_{15}$B$_{8}$] of the other selected cluster, [N$_{63}$B$_{44}$], contains the [N$_{13}$B$_{4}$] fragment, the two nnn N atoms along the $C_{2x}$ axis plus the four borons bridging those two nnn nitrogens with the [N$_{13}$B$_{4}$] subunit.
The calculations are carried out using the 6-41G basis set for B and the 7-311G basis for N \footnote{Removing the polarization $d$ function has negligible effects ($<<$1\%) on the on-site, nn, and nnn orbital relaxation contributions.}. 

The overall correlation correction associated with orbital relaxation and polarization effects at the six nnn nitrogens is listed on the fourth line of Table \ref{diagonalhole}. It amounts to --0.14 eV and was obtained as a sum of the correlation corrections originating from the orbital relaxation and polarization at the four nnn N in [N$_{71}$B$_{36}$] and the two nnn N in [N$_{63}$B$_{44}$]. To confirm that those corrections are indeed additive, we performed additional calculations on [N$_{71}$B$_{36}$], in which we allowed for the orbitals of only two of the four nnn N sites to relax and polarize. The relaxation effect for these two N atoms amounts to approximately --0.045 eV, which represents indeed one third of that --0.14 eV mentioned above. 
It is interesting to note that the correlation corrections associated with the nnn N sites are independent of the angular momentum of the orbital where the hole resides. This finding indicates that the interaction between the extra charge, placed at the central N, and the induced dipole at those nnn N sites is a monopole-dipole type interaction.

At last, we calculated the correction due to the relaxation of the orbital where the hole resides (fifth line in Table \ref{diagonalhole}). We accounted in this manner for the spin degree of freedom after the electron-removal. The hole orbital localized at the central N is now allowed to relax in the presence of nearby valence orbitals, on-site, nn, and nnn, that were allowed to relax in the first set of calculations.
The nearby nitrogen orbitals around the hole are kept frozen during this optimization of the singly occupied orbital.   

Worthwhile to mention, calculations of the on-site orbital relaxation and polarization correction, $\Delta$H$^{relax}_{mm}$($\textbf{0}$), exploiting either [N$_{13}$B$_{28}$], [N$_{55}$B$_{28}$] or [N$_{63}$B$_{44}$] clusters indicate that this quantity is converged with respect to the cluster size \footnote{$|\Delta H^{relax}_{mm}(\textbf{0})|$ increases by only 2 \% when enlarging the [N$_{55}$B$_{28}$] cluster to [N$_{63}$B$_{44}$].}. Similar observations hold for the relaxation and polarization corrections associated with the nn N sites.

Having computed the relaxation and short-range polarization effects on the diagonal matrix elements for the ($N\!-\!1\!$) valence-band states, we discuss next the low energy ($N\!+\!1\!$) conduction-band states. These electron-addition states imply  configurations with an extra electron into the B $2s$-like or B $2p$-like conduction-band WO's.  
We employed a [B$_{31}$N$_{68}$] cluster to investigate such configurations. 
The active region [BN$_{16}$] of the [B$_{31}$N$_{68}$] cluster consists of a central B site and the first two nitrogen coordination shells, containing four and twelve N atoms, respectively. The buffer region includes again all B and N sites within the first two coordination shells of the outer 12 active N atoms. We used in these calculations the 6-41G basis set for B and the 7-311G basis set for N.

In analogy to the ($N\!-\!1\!$) valence-band states, we have carried out ROHF calculations for the lowest-energy ($N\!+\!1\!$) states keeping the singly-occupied B $2s$-like or B $2p$-like orbitals frozen \cite{Pahl}. The valence $2s$ and $2p$ orbitals of the N atoms within the active region [BN$_{16}$] are allowed to relax and polarize in response to the electron-addition process. 
 \begin {table}
 \centering 
 \caption{Correlation-induced corrections (in eV) to the diagonal Hamiltonian matrix elements for the conduction-band B $2s$ and B $2p$ electron-addition states. Negative corrections indicate shifts to lower energies of the conduction bands.}
\begin{tabular}{p{5cm}p{1.6cm}p{1.6cm}}
  \hline\hline
 &$\Delta$ H$_{nn}$($\textbf{0}$)& \\ \hline 
&B $2s$-like&B $2p$-like\\ \hline
nn(N) orb. relaxation&-1.03&-1.14\\
nnn(N) orb. relaxation & -0.42 & -0.45 \\
Added-elec. orb. relaxation &-0.14& -0.18\\
Loss of ground-state corr. &0.60&0.50\\
Long-range polarization &-1.77& -1.77\\
Total corr. correction &-2.76&\textbf{-3.04}\\ 
 \hline \hline
 \end{tabular} 
\label{diagonalelec}
\end {table}

The on-site relaxation effect associated with the B $1s$ core orbital  is found to be vanishingly small. 
The relaxation of the nn nitrogen atoms gives rise to a downward shift of 1.0--1.1 eV for the B $2s$ and $2p$ conduction bands, see Table \ref{diagonalelec}. Relaxation effects at the nnn N sites bring an additional downward shift of about 0.45 eV of the center of gravity of the four low-lying conduction bands. 

In a next step, the B $2s$-like or B $2p$-like orbital with the extra electron is also allowed to relax, while keeping the other valence and low-lying conduction-band orbitals frozen. The reason for freezing the latter, i.e., the unoccupied B $2s$ and $2p$-like components, is to prevent delocalization of the extra electron.
For the same reason, in the case of the B $2p$ added-electron states, the next four higher-energy virtual orbitals are kept frozen as well.
Those are linear combinations of $s$ and $p$ virtual orbitals centered at the nearest N and B sites. 
Worthwhile to note is the similar magnitude of this correction as compared to the relaxation correction found for the N $2s, 2p$ hole orbitals (see Table \ref{diagonalelec}).  

An analysis of basis set effects similar to that performed for the ($N\!-\!1\!$) states was also carried out for the electron-addition states. Additional calculations of the relaxation effects associated with the nn N atoms were performed on a [B$_{13}$N$_{28}$] cluster. 
The active region of this cluster contains the central B site and the four nn N sites. The second nitrogen coordination shell of the central B (i.e., nnn N atoms) is  now included in the buffer region.
ROHF wave functions were constructed for the B $2s$ and $2p$ ($N\!+\!1\!$) states  using for the boron and nitrogen atoms either B 6-41G and N 7-311G or B 6-311G* and N 7-311G* basis sets. 
As compared to the ($N\!-\!1\!$) states, the effect is somewhat larger for the ($N\!+\!1\!$) states, 6 to 7 \%. 
The correlation-induced correction associated with the short-range relaxation at the four nn N sites (first line in Table \ref{diagonalelec}) includes this contribution of 6-7 \% related to basis set effects.

To summarize this subsection, the on-site orbital relaxation and short-range relaxation and polarization effects at the nn and nnn nitrogen sites in response to the extra charge, hole or electron, give rise to corrections of several eV to the HF band gap. 
\subsubsection{Long-range polarization effects}
A substantial contribution to the reduction of the HF gap also arises from long-range polarization effects. A simple estimate of the long-range polarization contribution to the shift of the energy bands is assessable through the computation of the classical polarization energy of a dielectric medium outside a sphere of radius $R$, beyond which the dielectric response of the crystal reaches its asymptotic value, $\epsilon_0$  \cite{Fuldebook}:
 \begin{eqnarray}
\triangle E (R)=-\frac{\epsilon_0-1}{2\epsilon_{0}}\frac{e^{2}}{R}
\nonumber
 \end{eqnarray}
This approximation is applicable because the interaction of the extra charge with the surroundings beyond a given $R$ has predominantly electrostatic character. The relaxation and short-range polarization effects \textit{within} the sphere were already computed at the \textit{ab initio} level and are summarized in Tables \ref{diagonalhole} and \ref{diagonalelec}.   

Using the experimental static dielectric constant, $\epsilon_0$=6.8 \cite{DC}, we find corrections of few eV to the diagonal matrix elements for the valence- and conduction-band states. Those values are also included in Tables \ref{diagonalhole} and \ref{diagonalelec}. 
The cuttoff radii $R$ for the ($N\!-\!1\!$) and ($N\!+\!1\!$) excited states, are each obtained as the average of the radii of the second and third nitrogen coordination shells around the central N or B sites. The corresponding values are 4.021 \AA $\,$ for the valence-band hole states and 3.468 \AA $\,$ for the conduction-band electron-addition states. 
The long-range polarization corrections $\Delta H^{lr}_{mm (nn)}(\mathbf{0})$ are then --1.53 eV for the N $2s$, $2p$ hole states and --1.77 eV for the B $2s$, $2p$ electron-addition states. 

It is instructive to compare the \textit{ab initio} result for the polarization effect at the nnn N sites with that obtained within the dielectric continuum approximation. We recall that around each N site there are 12 N atoms in the first nitrogen coordination shell (nn N) and 6 N sites in the second N coordination shell (nnn N). 
The \textit{ab initio} value for the contribution of the nnn N atoms to the overall correlation-induced correction to the diagonal matrix elements for the N $2s$, $2p$ hole states was found to be --0.14 eV, see Table \ref{diagonalhole} and the paragraphs above. 
A separate estimate for this quantity can be obtained within the dielectric continuum approximation. In this approximation, the correction to the diagonal matrix elements due to polarization effects at the second shell of N neighbors is $\delta E_{2}=S_{2}\frac{\epsilon_0-1}{2\epsilon_{0}}e^{2}(\frac{1}{R_1}-\frac{1}{R_2})$, with $R_1$=3.086 \AA, $R_2$=4.021 \AA $\,$, and $S_{2}$ =0.47. 
Here, $S_{2}$ is a scaling factor which represents the ratio between the density of N sites in the spherical shell enclosed between $R_1$ and $R_2$ and the density of N atoms in the unit cell, see also \cite{Hozoi}. The correction $\delta E_{2}$ is found to be --0.21 eV, which is about $30 \%$ larger than the \textit{ab initio} value of --0.14 eV. 
This difference between the {\it ab initio} estimate and the value deduced on the basis of the continuum dielectric model is not very surprising.
A better agreement between the two values would imply that a continuum approximation may be already  made for the polarization beyond the \textit{first} shell of N neighbors. This can not be expected to hold because at such short distances the associated charge distribution and polarization is not that of a continuum. Therefore, 
the errors related to the continuum dielectric approximation are larger for small $R$'s. 

\subsubsection{Loss of ground-state correlation}
An important contribution to the correlation-induced corrections to the diagonal matrix elements also arises from differential correlation effects, i.e., loss of ground-state correlation. 
To study such effects, we construct the correlated wave functions $|\Psi^{N-1}_{\mathbf{R_{J}}m\sigma'}\rangle$ and $|\Psi^{N+1}_{\mathbf{R_{I}}n \sigma}\rangle$ by means of SDCI calculations. 

The localized character of the valence-band WO's allows for the use of a [N$_{13}$B$_{28}$] cluster, with a single N atom in the active region [NB$_{4}$], for studying correlation effects beyond charge relaxation and polarization. The basis sets exploited in the SDCI calculations are 
 6-311G* for the B and 7-311G* for the N atoms. 

The SDCI wave functions for the $N$ and ($N\!-\!1\!$) electron states are constructed by correlating the $2s$ and $2p$ orbitals of the central N atom in the [NB$_{4}$] kernel. For the ($N\!-\!1\!$) electron configuration, the reference wave functions for the different N $2s$ and N $2p$ hole states $|\breve{\Phi}^{N-1}_{\mathbf{R_{J}}m\sigma'}\rangle$ are each expressed in terms of individually optimized orbital sets. The orbital with the hole is also relaxed, as described above.
In the framework of the quasiparticle approximation, the SDCI wave function for the ($N\!-\!1\!$) state should be expanded in terms of configuration state functions (CSF's) for which the occupation of the N $2s$ or $2p$ hole spin-orbital is kept frozen. This is referred to as
 the frozen local hole approximation \cite{Pahl}.
In practice, we work with spatial orbital rather than spin-orbital sets. In \textsc{molpro}, for example, the configuration selection is done for spatial orbitals and hence the SDCI wave function, constructed by correlating the singly occupied N $2s$ or $2p$ orbital with the doubly occupied N $2s$, $2p$ orbitals, may contain CSF's for which the occupation of the hole orbital is no longer maintained.    
The contributions of such CSF's to the total correlation energy for the ($N\!-\!1\!$) state account for effects beyond the loss of ground-state correlation and the quasiparticle picture. The associated excitations account for satellite structures in the photoionization spectra.
To assess the various correlation contributions, in particular, the loss of ground-state correlation, we analyzed the correlation energy for the ($N\!-\!1\!$) states in terms of intra- and inter-pair contributions of the N $2s$ and $2p$ orbitals. 

Consider one of the three N $2p$ hole states. In a first step, we included in the correlation treatment all singlet and triplet electron pairs excited from the doubly occupied N $2s$ and $2p$ orbitals and the singly occupied N $2p$ hole orbital.
The associated SDCI correlation energy of about --4.00 eV is partitioned into contributions of excited internal, singly external, and doubly external configurations. The terminology used for the different types of configurations follows that adopted by Werner and Knowles in Ref.\ ~\onlinecite{ICCI}. 
The excited singly external configurations are related to one-particle and semi-external two-particle excitations for which one electron is excited from a singly or doubly occupied orbital to a virtual (external) orbital. The excited doubly external configurations result from two-particle excitations for which both electrons are promoted to the virtual orbital space, i.e., external two-particle excitations.

We analyzed the different contributions of these excited internal, singly external, and doubly external configurations to the total SDCI correlation energy. 
With the present variational orbital space, excited doubly external configurations do not change the occupation of the hole orbital. This means that the intra- and inter-pair correlation contributions of the N $2s$ and $2p$ orbitals originating from such configurations are a relevant part in the computation of the loss of ground-state correlation.
Contrary to the doubly external configurations, the excited internal and the singly external configurations in which the occupation of the hole orbital is altered account for effects beyond the quasiparticle picture and frozen local hole approximation. Hence, the associated correlation contributions should not be considered in the evaluation of the loss of ground-state correlation. 
Our calculations indicate that with the present choice of the cluster, the total SDCI correlation energy of --4.00 eV contains a negligibly small contribution from excited internal configurations, less than --0.001 eV. 
The overall correlation contribution arising from the excited singly external configurations is --0.53 eV.   
We are interested in determining the fraction due to configurations in which the occupation of the hole orbital is preserved. 
An analysis of the SDCI wave function indicate comparable coefficients for the excited singly external CSF's in which the N $2p$ hole orbital is either doubly or singly occupied.
The configurations with a doubly occupied N $2p$ ``hole"  orbital arise from so-called semi-external two-particle excitations. 
Since the composition of the SDCI wave function is not sufficient to determine the relevant part of the net effect, we carried out further the following analysis.

In a second step, we computed all various intra- and inter-pair contributions of the N $2s$ and $2p$ orbitals by correlating in separate SDCI calculations different orbitals and orbital pairs, respectively. 
We constructed thus SDCI wave functions where either one orbital, a pair of orbitals, or a combination of any three orbitals from the N $2s$, $2p$ orbital set were correlated. 
These wave functions were carefully examined to determine the singly external CSF's with large coefficients in the CI expansions. 
We found that the correlation energy due to excited singly external configurations in which the occupation of the N $2p$ hole orbital is unaltered is about --0.25 eV. 
The remaining --0.28 eV are associated with configurations with a double occupation of the N $2p$ ``hole" orbital. 
Worthwhile to mention, the inter-pair correlation contribution involving the N $2p$ hole orbital and the remaining, doubly occupied N $2s$, $2p$ orbitals arises from external two-particle excitations, i.e., excited doubly external configurations. This contribution amounts to about --0.84 eV. 

The findings above indicate that the total correlation energy of --4.00 eV contains a moderate contribution of about --0.28 eV originating from configurations in which the occupation of the hole orbital is no longer maintained. 
In order to obtain the loss of ground-state correlation, the difference of --3.72 eV should be compared with the correlation energy for the $N$-particle ground state.
The correlation energy for the $N$-particle ground state is found to be about --4.77 eV. Hence, the correction to the diagonal matrix elements, $\Delta H^{lgsc}_{mm}(\mathbf{0})$, due to the loss of ground-state correlation shifts the N $2p$ bands downward by about 1.05 eV; see Table \ref{diagonalhole}. 

The loss of ground-state correlation for the N $2s$ hole state is more difficult to calculate because maintaining the hole at the N $2s$ orbital in all configurations of the SDCI wave function is technically not possible. 
 We calculated the associated correction by using only the correlation energy contribution of the excited doubly external configurations in the wave function $|\Psi^{N-1}_{\mathbf{R_{J}}m\sigma'}\rangle$ of the N $2s$ hole state. 
The energy contribution arising from excited singly external configurations is predominantly due to semi-external two-particle excitations resulting into a double occupation of the N $2s$ orbital. Hence, this contribution is beyond the loss of ground-state correlation. With the current choice of the cluster, no internal configurations contribute to the net correlation energy of the N $2s$ hole state. 
Our result is that the overall correlation correction shifts downward the center of gravity of the N $2s$ bands by about 1.08 eV. 

Similar analyses were also carried out for the conduction-band ($N\!+\!1\!$) states. The cluster exploited in these studies is [B$_{13}$N$_{28}$], with an active region [BN$_{4}$]. The basis sets employed are 6-311G* for the B atoms and 7-311G* for N. The SDCI wave functions for the ($N\!+\!1\!$) electron states are constructed by correlating explicitly the $2s$ and $2p$ orbitals of the four N atoms around the central B site. 
The reference wave functions in the correlation treatment are the relaxed $(N+1)$ wave functions $|\breve{\Phi}^{N+1}_{\mathbf{R_{I}}n\sigma}\rangle$. 

Given the configuration selection in \textsc{molpro}, in a quasiparticle picture, the correlation between the doubly occupied N $2s$ and $2p$ orbitals and the singly occupied B $2s$-like or $2p$-like orbital can be accounted for by SDCI wave functions where only the doubly occupied N $2s$ and $2p$ orbitals are correlated. 
The singly occupied B $2s$-like or $2p$-like orbital is placed in the active orbital space, but no excitations out of this orbital are included in the correlation treatment.  
Such a SDCI wave function incorporates correlation contributions of all relevant internal, excited singly external, and doubly external configurations. 
The selected excitation space contains thus internal excitations and semi-external two-particle excitations which promote an electron from a doubly occupied N $2s$ or $2p$ orbital to the B $2s$ or $2p$ orbital and a second electron from the doubly occupied to the virtual orbital space.
External one-particle and two-particle excitations which involve electrons or electron pairs from the doubly occupied N $2s$ and $2p$ orbitals are also included in the excitation domain.  
In analogy with the considerations for the ($N\!-\!1\!$) states, we compared the total correlation energy for the ($N\!+\!1\!$) state, corresponding to SDCI wave functions of the type described above, with the correlation energy for the $N$-particle ground state. 
The SDCI wave function for the $N$-particle configuration is constructed by correlating the doubly occupied N $2s$ and $2p$ orbitals of the four N atoms around the central B site.
The vacant B $2s$-like and $2p$-like orbitals are all part of the variational orbital space in this case.

 The corrections to the diagonal matrix elements, $\Delta H^{lgsc}_{nn}(\mathbf{0})$, due to the loss of ground-state correlation are similar for the B $2s$ and $2p$ electron-addition states, about 0.5 eV. The results are listed in Table \ref{diagonalelec}.

Although beyond the scope of this study, we mention that SDCI wave functions containing configurations in which the added electron is not maintained in the same B $2s$ or $2p$ orbital as in the reference wave function are related to satellite structures in the inverse photoionization spectra. 
These configurations describe correlation effects beyond the loss of ground-state correlation.
Such SDCI wave functions can be constructed by including in the correlation treatment not only the doubly occupied N $2s$ and $2p$ orbitals, but also the singly occupied B $2s$-like or $2p$-like orbital. 
The inclusion of the singly occupied B $2s$ or $2p$ orbital in the correlation treatment leads to an additional contribution of about --1.0 eV.  The latter effect is predominantly due to excited doubly external configurations involving the extra electron.
Such differential correlation effects lead thus to a stabilization of the ($N\!+\!1\!$) electron configuration with respect to the $N$-particle ground state by --0.5 eV. 

At this point, we have computed all relevant correlation-induced corrections to the diagonal matrix elements for the valence- and conduction-band states. 
The overall effect is a reduction of the HF band gap of $c$-BN from 13.62 eV to 6.91 eV. 
This result compares well with experimental estimates of 6.0$\pm$0.5 eV \cite{Fomichev, Agui} and 6.4$\pm$0.5 eV \cite{Chrenko}. 
More sophisticated basis sets are not expected to affect significantly the calculated values, since the present choice of basis sets proves to be quite reasonable. For comparison, LDA calculations carried out with the triple-zeta GTO basis set used in the HF calculations yield a too small band gap of 4.34 eV, in agreement with other LDA studies; see, e.g., Refs. ~\onlinecite{Renata, Park}. An exception makes the LDA+$GW$ study from Ref. ~\onlinecite{Surh}, which provides a theoretical value of 6.3 eV for the indirect band gap of c-BN.
\subsection{Off-diagonal matrix elements of the local Hamiltonian}
We discuss next the correlation-induced corrections to the band widths. The computation of these corrections requires the off-diagonal matrix elements of the effective Hamiltonian in Eq. (\ref{corbands}).

We consider first the Hamiltonian matrix elements between two frozen-orbital ROHF wave functions $|\Phi^{N-1}_{\mathbf{R_{J}}m\sigma'}\rangle$ having the hole at distinct sites $\mathbf{R_{J}}$, see Eqs. (\ref{rsf}, \ref{LHSCF2}). 
The off-diagonal Hamiltonian matrix elements $H^{\textsc{scf}}_{\mathbf{R_{J}}, mm'}$ between such mutually orthogonal wave functions $|\Phi^{N-1}_{\mathbf{R_{J}}m\sigma'}\rangle$ and $|\Phi^{N-1}_{\mathbf{0}m'\sigma'}\rangle$ constitute the so-called hopping terms in an orthogonal tight-binding approach.
The same considerations hold for the wave functions  $|\Phi^{N+1}_{\mathbf{R_{I}}n\sigma}\rangle$ and Hamiltonian matrix elements $H^{\textsc{scf}}_{\mathbf{R_{I}}, nn'}$.    

In a second step, we incorporate relaxation and polarization effects in the nearby surroundings of the hole or extra electron. This is achieved by separate SCF optimizations for each of the ($N\!-\!1\!$) and ($N\!+\!1\!$) electron states. Hence, we construct the correlated wave functions $|\tilde{\Phi}^{N-1}_{\mathbf{R_{J}}m\sigma'}\rangle$, $|\tilde{\Phi}^{N-1}_{\mathbf{0}m'\sigma'}\rangle$ and $|\tilde{\Phi}^{N+1}_{\mathbf{R_{I}}n\sigma}\rangle$, $|\tilde{\Phi}^{N+1}_{\mathbf{0}n'\sigma}\rangle$. The separate SCF optimizations lead to sets of non-orthogonal orbitals. 
A scheme for the computation of the Hamiltonian and overlap matrix elements between such non-orthogonal wave functions was recently implemented in {\sc molpro} by Mitrushchenkov and Werner \cite{Mitrushchenkov}. It is based on non-unitary transformations of the initial set of non-orthogonal orbitals to biorthogonal sets and follows closely the approach proposed by Malmqvist \cite{Malmqvist}.
Alternative approaches were proposed in other groups, see, e.g., Ref.~\cite{Broer}.  

There are two different routes into constructing the correlated energy bands $\epsilon_{\mathbf{k'}\mu\sigma'}$ and $\epsilon_{\mathbf{k}\nu\sigma}$. The first possibility is to diagonalize a $\mathbf{k}$-dependent matrix like that in Eq.\ (\ref{corbands}), which contains the Hamiltonian and overlap matrix elements between the non-orthogonal, correlated wave functions $|\tilde{\Phi}_{\mathbf{R_{I(J)}}}^{N\pm1}\rangle$ or $|\Psi_{\mathbf{R_{I (J)}}}^{N\pm1}\rangle$. 
A second route rests on deriving from the initial inter-site Hamiltonian and overlap matrix elements a set of effective hopping integrals for an orthogonal tight-binding-like formulation. In the latter approach, the effective hopping terms can be directly compared with the HF hoppings, which offer a more intuitive picture onto the effect of correlations on the electronic band structure.   

For electron-removal states which have different binding energies and are non-orthogonal, i.e., $H_{\mathbf{0}, m'm'} \neq H_{\mathbf{R_J}, mm}$ and $S_{\mathbf{R_J}, mm'}\neq 0$, the effective hopping matrix elements are defined as
 \begin{eqnarray}
 \nonumber
 t_{mm'} (\mathbf{R_{J}})&=&\frac{1}{1-S^{2}_{\mathbf{R_{J}}, mm'}}\Bigg\{H_{\mathbf{R_{J}}, mm'}-
 \nonumber   \\
 &-&S^{2}_{\mathbf{R_{J}}, mm'}\bigg\{\frac{H_{\mathbf{0}, m'm'}+H_{\mathbf{R_J}, mm}}{2}\bigg\}\Bigg\}
 \nonumber \\
 &=& \frac{1}{2}\Bigg\{\triangle E^{2}-\frac{(H_{\mathbf{R_{J}}, mm}-H_{\mathbf{0}, m'm'})^{2}}{1-S^{2}_{\mathbf{R_{J}}, mm'}}\Bigg\}^{\frac{1}{2}}
 \nonumber
\end{eqnarray}
and analogously for  $t_{nn'} (\mathbf{R_{I}})$. Here, $\triangle E$ is the energy separation between the two eigenstates of the 2 x 2 CI secular problem, 
$|H-ES|=0$. This type of secular problem in terms of non-orthogonal sets of orbitals will be referred to as non-orthogonal CI (NOCI); see, e.g., \cite{Broer}. 
 If the two wave functions are orthogonal, $\langle\tilde{\Phi}_{\mathbf{0}m'\sigma'}^{N-1}|\tilde{\Phi}_{\mathbf{R_J}m\sigma'}^{N-1}\rangle$=0, the effective hopping term reduces to
\begin{eqnarray}
 \nonumber
 t_{mm'} (\mathbf{R_{J}})&=&H_{\mathbf{R_{J}}, mm'} 
 \nonumber \\
 &=&\frac{1}{2}\Bigg\{\triangle E^{2}-(H_{\mathbf{R_{J}}, mm}-H_{\mathbf{0}, m'm'})^{2}\Bigg\}^{\frac{1}{2}}.
 \nonumber
\end{eqnarray}

\begin{figure}[htbp]
\begin{center}
\includegraphics[width=8.5cm]{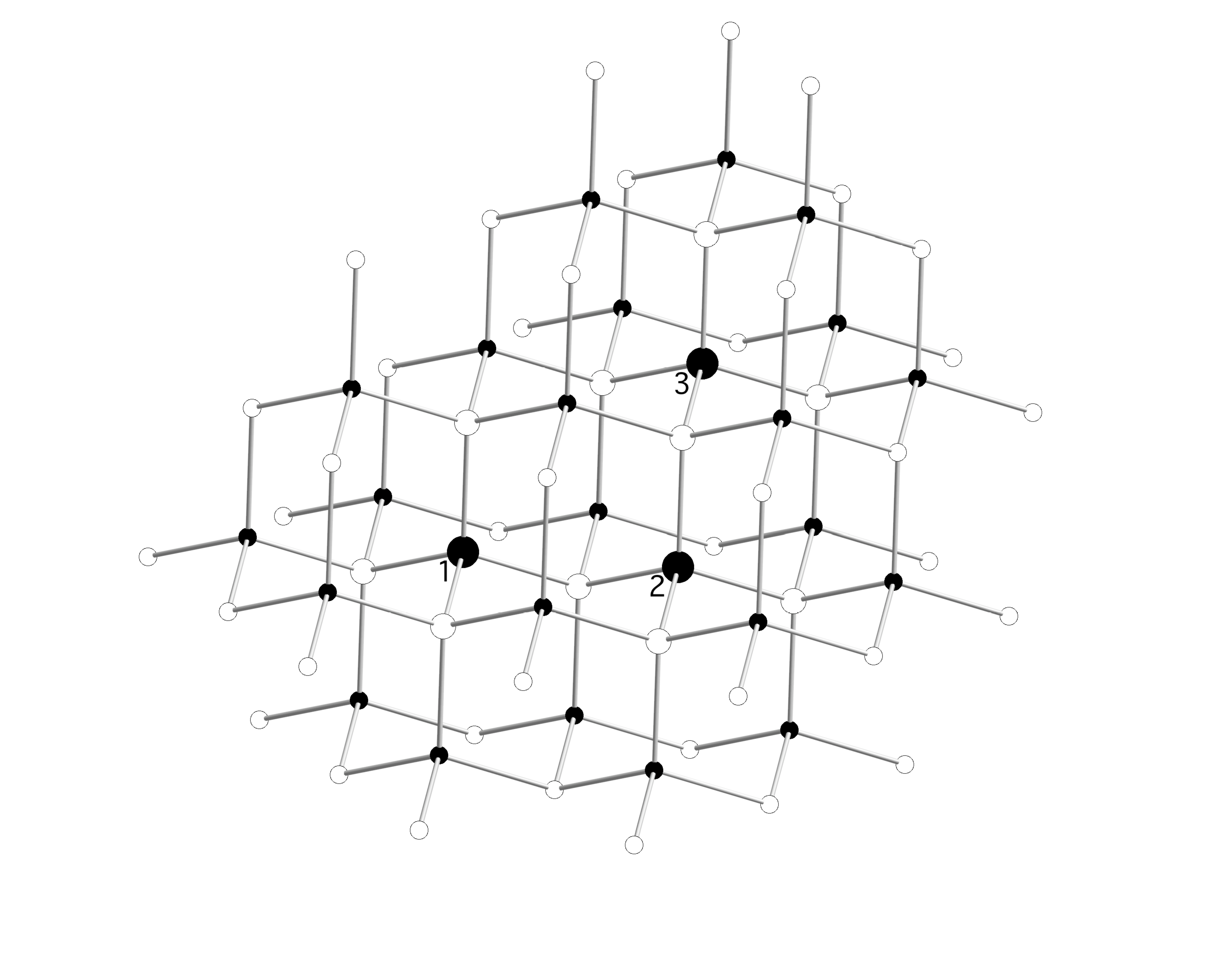}
\caption{Illustration of the [N$_{26}$B$_{50}$] embedded cluster used for the computation of the nnn effective hopping terms associated with the valence-band N $2s$, $2p$ hole states. The active region of the cluster consists of the [N$_{3}$B$_{10}$] fragment. The three N sites in the active region are depicted as large black spheres and labeled as 1, 2, 3. N atoms in the buffer zone are shown as small black spheres. The ten B atoms in the active region are drawn as large white spheres, whereas the buffer B atoms are shown as smaller white spheres. }
\label{clusterhop}
\end{center}
\end{figure}

To compute the nn and nnn effective hoppings associated with the N $2s$, $2p$ valence bands, we designed two different clusters, [N$_{20}$B$_{40}$] and [N$_{26}$B$_{50}$]. 
The active regions of these clusters are the [N$_{2}$B$_{7}$] and [N$_{3}$B$_{10}$] fragments, respectively. 
The [N$_{2}$B$_{7}$] fragment consists of two N sites, denoted as 1 and 2 in Fig. \ref{clusterhop}, plus their nn B atoms. Likewise, the [N$_{3}$B$_{10}$] active region contains the N sites 1 and 2 plus the N atom denoted as 3 in Fig. \ref{clusterhop}. The nn B atoms of the three N sites are also included in the active subunit. The buffer regions of the two clusters consist of all N and B atoms in the first two coordination shells of the active boron sites. 
The [N$_{26}$B$_{50}$] cluster is sketched in Fig. \ref{clusterhop}.

The basis sets exploited in the calculations for the [N$_{20}$B$_{40}$] cluster are 7-311G* for the N atoms and 6-311G* for B. 
In the case of the larger [N$_{26}$B$_{50}$] cluster, we used the 6-41G basis set for the boron atoms since otherwise the computation becomes unfeasible.
In analogy with the analysis for the diagonal matrix elements, basis set effects for the effective hopping terms were checked by additional calculations for the nn hoppings using the [N$_{20}$B$_{40}$] cluster. In this extra set of calculations, the smaller B 6-41G basis set was used. 
We found negligibly small basis set effects for the nn hopping terms, of the order of 1\%.

An overview of the valence-band nn and nnn effective hoppings is provided in Table \ref{Hop1}. 
\begin {table}
 \centering 
 \caption{Nearest-neighbor (nn), $\mathbf{R_J}=(0, 1, -1)\frac{a}{2}$, and next-nearest-neighbor (nnn), $\mathbf{R_J}=(0, 1, 0) a$, effective hopping matrix elements for the valence-band states (in eV).  Frozen-orbital CI results (FOCI ) are given in the second column. NOCI matrix elements in terms of relaxed orbitals are listed in the third column (RO-NOCI). NOCI results for which the orbital where the hole resides is also relaxed are listed in the fourth column. Matrix elements smaller than 0.01 eV are not included in the table.}
\begin{tabular}{p{2.0cm}p{2.0cm}p{2.0cm}p{2.0cm}}
  \hline\hline
Active WO's &\multicolumn{3}{c}{nn t$_{mm'}$($\mathbf{R_J}$)} \\ 
& FOCI&RO-NOCI&ROH-NOCI\\ \hline
$2s$-$2s$ & 0.520& 0.541& 0.544\\
$2p$$_x$-$2p$$_x$ & 0.184&0.183&0.187 \\
$2p$$_{y (z)} $-$2p$$_{y (z)} $& 0.724&0.739&0.747\\
$2p$$_{y (z)} $-$2p$$_{z (y)} $& 0.928 &0.954 & 0.967 \\
$2p$$_{x (y)} $-$2p$$_{y (x)} $&0.053 (0.063) &0.060 (0.070) & 0.062 (0.072)  \\
$2p$$_{x (z)} $-$2p$$_{z (x)} $&0.053 (0.063) &0.060 (0.070) & 0.062 (0.072) \\
\hline
  \hline
Active WO's &\multicolumn{3}{c}{nnn t$_{mm'}$($\mathbf{R_J}$)} \\ 
& FOCI&RO-NOCI&ROH-NOCI\\ \hline
$2s$-$2s$ &0.011& 0.012&0.012 \\
$2p$$_x$-$2p$$_x$ &0.054 &0.050&0.049 \\
$2p$$_{y} $-$2p$$_{y} $&0.037 &0.030&0.028\\
$2p$$_{y (z)} $-$2p$$_{z (y)} $& 0.053 & 0.050&  0.072\\
\hline\hline
\end{tabular} 
\label{Hop1}
\end {table}
We list in the first column the Wannier functions where the hole resides.
In the second column, we include results of frozen-orbital CI calculations in terms of the initial HF WO's (FOCI). 
In a next step, we allow for full relaxation of the doubly occupied $2s$ and $2p$ orbitals of the N atoms within the active regions (RO-NOCI). 
The orbital optimizations were carried out separately for each electron-removal wave function $|\Phi^{N-1}_{\mathbf{R_{J}}m\sigma'}\rangle$ and $|\Phi^{N-1}_{\mathbf{0}m'\sigma'}\rangle$, yielding the wave functions $|\tilde{\Phi}^{N-1}_{\mathbf{R_{J}}m\sigma'}\rangle$ and $|\tilde{\Phi}^{N-1}_{\mathbf{0}m'\sigma'}\rangle$, respectively. The corresponding effective hoppings are listed in the third column.
Finally, results where in addition to the relaxation of the N $2s$, $2p$ doubly occupied orbitals, we also relax the hole orbital are given in the fourth column of Table \ref{Hop1} (ROH-NOCI).  

The results indicate small changes of about 0.02 eV for the nn effective hopping terms when relaxation effects in the nearby surroundings are accounted for.
The general trend is a slight increase of the nn hoppings t$_{mm'}$, which persists further upon the relaxation of the hole orbital. 
In the latter case, the changes are also of the order of 0.01 eV.%

Owing to the localized nature of the valence-band hole states, the effective hopping matrix elements decay fast with the distance $\mathbf{R_J}$. 
The nnn hopping terms are found to be an order of magnitude smaller than the nn matrix elements. The short-range relaxation and polarization corrections for the nnn terms are also an order of magnitude smaller as compared to those for the nn elements, see Table \ref{Hop1}. 

Some trial calculations were also performed to determine the role of
correlation effects beyond relaxation and polarization on the hopping matrix elements.   
As for the case of diagonal matrix elements, we employed CI wave functions with single and double excitations from the relaxed 
$2s$ and $2p$ orbitals of the active N sites.  We found that the SDCI treatment leads only to a slight reduction of the effective valence-band hoppings, with variations of up to 0.02 eV. 
Since these additional corrections are very small and the details regarding the construction of CI expansions where correlation effects beyond the loss of ground-state correlation are minimized require very technical explanations, we do not discuss these effects any further. 

Summarizing our results for the effect of correlations on the nn and nnn intersite interactions, we find that the widths of the upper-lying valence bands are only little affected. 
The fact that the hopping matrix elements from the calculations in terms of relaxed orbitals are slightly enlarged as compared to the FOCI values is due to the larger intersite overlap between the relaxed orbitals of the two wave functions
$|\tilde{\Phi}^{N-1}_{\mathbf{R_{J}}m\sigma'}\rangle$ and $|\tilde{\Phi}^{N-1}_{\mathbf{0}m'\sigma'}\rangle$. 
The N $2p$ orbitals within the nn region of the N site where the hole resides are polarized toward the positive charge at this site. It is this orbital polarization effect that leads to a larger intersite orbital overlap and enhanced hoppings.
Similar polarization effects were also discussed for the case of O hole states in MgO \cite{Hozoi}.  %
We have not considered explicitly the effect of long-range polarization on the off-diagonal matrix elements. This effect, however, is expected to be negligible.  

A similar analysis was carried out for the lowest electron-addition states and conduction-band hoppings. The clusters employed to investigate the nn and nnn  B-B interactions are designed in perfect analogy with the [N$_{20}$B$_{40}$] and [N$_{26}$B$_{50}$] clusters described above. 
The active regions of these [B$_{20}$N$_{40}$] and [B$_{26}$N$_{50}$] clusters are the kernels [B$_{2}$N$_{7}$] and [B$_{3}$N$_{10}$], respectively. 
Results of both FOCI and NOCI calculations are summarized in Table \ref{Hop2}. In the NOCI calculations, the $2s$ and $2p$ orbitals of the nearest N atoms are allowed to relax in response to the extra electron in the conduction-band B $2s$ or $2p$ orbital.   
\begin {table}
 \centering 
 \caption{Nearest-neighbor (nn), $\mathbf{R_I}=(0, -1, 1)\frac{a}{2}$, and next-nearest-neighbor (nnn), $\mathbf{R_I}=(0, -1, 0) a$, effective hopping matrix elements for the conduction-band states (in eV). FOCI results are listed in the second column. The third column summarizes the NOCI data. Matrix elements smaller than 0.01 eV are not included in the table.}
\begin{tabular}{p{2.8cm}p{2.2cm}p{2.1cm}}
  \hline\hline
Active WO's &\multicolumn{2}{c}{nn t$_{nn'}$($\mathbf{R_I}$)} \\ 
&FOCI&RO-NOCI\\ \hline
$2s$-$2s$ &0.040 &0.061 \\
$2p$$_x$-$2p$$_x$ &1.571 &1.626 \\
$2p$$_{y (z)}$-$2p$$_{y (z)}$&0.383& 0.377 \\
$2p$$_{x (y)}$-$2p$$_{y (x)}$&0.269 &0.255 \\
$2p$$_{x (z)}$-$2p$$_{z (x)}$&0.269 &0.255 \\
$2p$$_{y(z)} $-$2p$$_{z (y)}$&0.168 &0.190 \\
\hline
  \hline
Active WO's &\multicolumn{2}{c}{nnn t$_{nn'}$($\mathbf{R_I}$)} \\ 
& FOCI&RO-NOCI\\ \hline
$2p$$_x$-$2p$$_x$& 0.177&0.164 \\
$2p$$_z$-$2p$$_z$&0.131 &0.116 \\
$2p$$_{y}$-$2p$$_{y}$&0.070&0.063  \\
$2p$$_{y (x)}$-$2p$$_{x (y)}$&0.031 & 0.030\\
$2p$$_{x (z)}$-$2p$$_{z (x)}$&0.062 &0.060 \\
\hline\hline
\end{tabular} 
\label{Hop2}
\end {table}
The additional effects resulting from the relaxation of the orbital where the extra electron resides are difficult to estimate for the nn hopping terms associated with the ($N+1$) states because the orbital could not be restrained from delocalizing over the two B sites in the $(N+1)$ SCF calculation. 
Such effects could be assessed, however, for the nnn hoppings t$_{nn'}$ and proved to be of the same order of magnitude as that observed for the valence-band ($N\!-\!1\!$) states, i.e., of the order of 0.01 eV. 

The larger extent of the conduction-band WO's manifests itself in the larger values of the nn and nnn effective hopping terms and their somewhat slower decay 
with the distance $\mathbf{R_{I}}$. 
The orbital relaxation and short-range polarization effects on these matrix elements, though, remain relatively small, with corrections of about 0.02 eV. The largest change is observed for the $2p_x$-$2p_x$ intersite interaction, about 0.06 eV. 
Our results indicate modifications of the hopping terms in two directions: some matrix elements are slightly enlarged by taking into account the short-range relaxation and polarization, others are reduced. 
Similar findings were reported for the effective hoppings associated with the lower-lying Mg $3s$ and $3p$ conduction bands of MgO\cite{Hozoi}.

For the computation of the nnn hopping terms, the basis set employed for the B atoms is the smaller 6-41G set. Basis set effects were assessed by extra calculations for the nn hoppings, using either 6-41G or 6-311G* basis sets for the boron species. The cluster employed in these calculations is the [B$_{20}$N$_{40}$] cluster. 
With the 6-41G basis set, we find changes of 0.01 to 0.06 eV for the nn B $2p$-$2p$ hoppings and changes of about 0.05 eV for the nn B $2s$-$2s$ hoppings, both at the frozen-orbital HF and correlated levels. Basis set effects of similar magnitude, not larger than few tens of meV, are expected for the nnn conduction-band hoppings. 

To conclude, the overall corrections for the intersite interactions, related to the formation of the short-range polarization cloud around the added electron, are rather small, in the range of few percents. A similar situation has been also observed for the conduction-band states of the ionic insulator MgO\cite{Hozoi}. 
The widths of the HF valence and conduction bands change thus little by incorporating short-range relaxation and polarization effects. In contrast with the relatively weakly correlated BN compound, the effective hopping matrix elements associated with electron-removal and electron-addition states in strongly correlated transition-metal compounds like the perovskite manganites are reduced considerably, by a factor of 4, due to the presence of strong magnetic interactions in these systems \cite{Paper1}. 
A strong reduction of the effective hopping terms was also found for the case of cuprate compounds \cite{Liviu}.

At last, we evaluate the effect of the correlation-induced corrections to the widths of the bands on the size of the band gap. For this purpose, we consider first the 
energy $\epsilon_{\mathbf{k}\mu}$ of the N $2p$ band states at the $\Gamma_{15}^{v}$ point. 
For the $p_x$ band, for example, $\epsilon_{\mathbf{0}\mu, x}$=$E^{(000)}_{x,x}+8t^{(110)}_{x,x}+4t^{(011)}_{x,x}+ O (nnn) + ...$ \cite{SlaterKoster}, where $E^{(000)}_{x,x}$ is a diagonal matrix element. 
For symmetry reasons, $t^{(110)}_{x,x}=t^{(011)}_{y,y}=t^{(011)}_{z,z}$. Taking into account corrections of about 0.02 eV for $t^{(110)}_{x,x}$ and 0.003 eV for $t^{(011)}_{x,x}$, we find that the N $2p$ bands are shifted upwards  by about 0.17 eV at the $\Gamma$ point. The band gap is reduced then by the same amount.  
For the B $2s, 2p$ conduction-band complex, the correlation-induced corrections to the B $2p$-$2p$ intersite matrix elements are most relevant at the X$_{1c}$ symmetry point.
The four by four secular problem factors out on account of symmetry at the 
X point, such that one finds the degenerate B $2p_y$ and $2p_z$ states and the lower $2s$ and $2p_x$ bands. The energy of the $2p_x$ band at the symmetry point X$_{1c}$ ($k_x=0,k_y=\pi,k_z=\pi$) is given by
$\epsilon_{\mathbf{k}\mu, x}$=$E^{(000)}_{x,x}+4t^{(011)}_{x,x}-8t^{(110)}_{x,x}-2t^{(010)}_{x,x}-2t^{(001)}_{x,x}+2t^{(100)}_{x,x} +...$ \cite{SlaterKoster}. 
From symmetry considerations, $t^{(110)}_{x,x}= t^{(011)}_{y,y}=t^{(011)}_{z,z}$, $t^{(001)}_{x,x}=t^{(010)}_{z,z}$, and $t^{(100)}_{x,x}=t^{(010)}_{y,y}$. Corrections of about 0.05 eV for $t^{(011)}_{x,x}$ and 0.01 eV for $t^{(110)}_{x,x}$ and for the nnn hopping terms lead to a shift to lower energy of the B $2p$ band at the X$_{1c}$ point and a further reduction of the gap by about 0.24 eV. 
Our final result for the fundamental gap of BN is thus 6.50 eV, which compares well with experimental estimates of 6.0$\pm$0.5 eV \cite{Fomichev, Agui} and 6.4$\pm$0.5 eV \cite{Chrenko}.
\section{Conclusions}
We investigated in the present work the correlation-induced corrections to the Hartree-Fock band structure of zinc-blende BN. A local Hamiltonian approach and quasiparticle approximation were employed for this purpose. 
An important advantage of our wave-function-based method is that it allows for a rigorous, \textit{ab initio} determination of the relevant correlation contributions to the gap and band widths of a crystalline material without compromising its infinite nature. 
The correlation contributions associated with the short-range relaxation and polarization around an extra hole or electron were explicitly computed. Moreover, correlation effects beyond relaxation and polarization, such as the loss of ground-state correlation, were also investigated. 
The long-range polarization of the crystal was treated within the dielectric continuum approximation. 

We found that the on-site and nearest-neighbor relaxation and polarization bring contributions of few eV to the correlation-induced corrections to the gap of BN. 
The long-range polarization caused by the extra hole or electron leads to an additional reduction of about 2.0 eV of the HF band gap. 
The net effect is a reduction of the HF gap from 13.62 eV to a value of 5.36 eV. 
If, in addition, we take into account correlation effects beyond relaxation and polarization, such as the loss of ground-state correlation, the band gap increases to 6.91 eV.  At last, an additional small reduction of 0.41 eV to a value of 6.50 eV is caused by correlation-induced variations of the widths of the bands.  
The computed band gap compares well with that deduced from soft x-ray experiments at the B and N $K$-edges \cite{Fomichev, Agui} and optical absorption measurements \cite{Chrenko}. As expected, the LDA-DFT calculations yield a too small band gap of 4.34 eV. 

While the HF gap is substantially modified by correlations, the widths of the bands are less sensitive to such effects. The inclusion of local correlations leads to a slight broadening of the N $2p$ bands. This broadening is attributed to the polarization of the nearest-neighbor N $2p$ orbitals toward the site where the hole resides.
\begin{acknowledgments}
We thank Professor Beate Paulus (Berlin) for valuable suggestions in the early stages of this work.
\end{acknowledgments}

\appendix*
\section{6-311G* and 6-41G basis sets for the B atoms.}
Tables \ref{BS1} and \ref{BS2} list the two basis sets for the B atoms designed for the computation of the correlation-induced corrections to the HF gap of c-BN.
\begin {table}[h]
 \centering 
 \caption{Exponents (in a.u.) and coefficients of the 6-311G* GTO basis set used for the B atoms.}
\begin{tabular}{p{1.2cm}p{1.1cm}p{2.0cm}p{2.0cm}p{1.7cm}}
  \hline\hline
& & &\multicolumn{2}{c}{Coefficient} \\ 
Shell&Type &Exponent & s&p\\ \hline 
2&sp &13.5055949 &0.1099474&0.0420302 \\
& & 3.0900933&0.8728330 &0.2359840\\
& &0.9128560 & 0.5149721&0.7318254\\
3&sp &0.4927046 &1.0&1.0 \\
4&sp &0.213 &1.0&1.0 \\   \hline\hline
\end{tabular} 
\label{BS1}
\end {table}

\begin {table}[h]
 \centering 
 \caption{Exponents (in a.u.) and coefficients of the B 6-41G GTO basis set.}
\begin{tabular}{p{1.2cm}p{1.0cm}p{2.0cm}p{2.0cm}p{1.7cm}}
  \hline\hline
& & &\multicolumn{2}{c}{Coefficient} \\ 
Shell&Type &Exponent & s&p\\ \hline 
2&sp &13.9313977 &0.0853578&0.0418012 \\
& &3.1396444&0.7038039 &0.2468341\\
& &0.9037473 & 0.7225717&0.7649465\\
& &0.4893080 & -3.1330026& 0.7122470 \\
3&sp &0.1691601&1.0&1.0 \\ \hline\hline
\end{tabular} 
\label{BS2}
\end {table}


\begin{thebibliography}{120}
 %
 \bibitem{Hoh} P.\ Hohenberg and W.\ Kohn, Phys.\ Rev.\ \textbf{136}, 864 (1964).
 %
 \bibitem{KS} W.\ Kohn and L.\ Sham,  Phys.\ Rev.\ B \textbf{140}, 1133 (1965).
%
 \bibitem{HF} D.\ R.\ Hartree, Proc.\ Cambridge Philos.\ Soc.\ \textbf{24}, 89 (1928); V.\ Fock, Z.\ Phys.\  \textbf{61}, 126 (1930).
 % 
 \bibitem{Sun}J.-Q.\ Sun and R.\ J.\ Bartlett, J.\ Chem.\ Phys.\ \textbf{104}, 8553 (1996).
 %
 \bibitem{Ayala}P.\ Y.\ Ayala, K.\ N.\ Kudin, and G.\ E.\ Scuseria, J.\ Chem.\ Phys.\ \textbf{115}, 9698 (2001).
 %
  \bibitem{Pisani} C.\ Pisani, M.\ Busso, G.\ Capecchi, S.\ Casassa, R.\ Dovesi, and L.\ Maschio, C.\ Zicovich-Wilson, and M.\ Sch\"utz, 
J.\ Chem.\ Phys.\  \textbf{122}, 094113 (2005).
%
%
\bibitem{Forner} W.\ F\"orner, R.\ Knab, J.\ \v{C}\'{\i}\v{z}ek, and J.\ Ladik, J.\ Chem.\ Phys.\ \textbf{106}, 10248 (1997).
%
\bibitem{Karin} K.\ Fink and V Staemmler, J.\ Chem.\ Phys.\ \textbf{103}, 2603 (1995).
%
\bibitem{Hirata2} S.\ Hirata, R.\ Podeszwa, M.\ Tobita, and R.\ J.\ Bartlett, J.\ Chem.\ Phys.\ \textbf{120}, 2581 (2004).
%
\bibitem{Stoll1} H. Stoll, Phys.\ Rev.\ B  \textbf{46}, 6700 (1992).
%
\bibitem{Shukla2} A.\ Shukla, M.\ Dolg, P.\ Fulde, and H.\ Stoll, 
Phys.\ Rev.\ B\ \textbf{57}, 1471 (1998).
%
\bibitem{Gr2} J.\ Gr\"afenstein, H.\ Stoll, and P.\ Fulde,  Phys.\ Rev.\ B \textbf{55}, 13588 (1997).
%
\bibitem{Albrecht} M.\ Albrecht, P. Fulde, and H.\ Stoll, Chem.\ Phys.\ Lett.\ \textbf{319}, 355 (2000).
%
\bibitem{Birken2} U.\ Birkenheuer, P.\ Fulde, and H.\ Stoll, Theor.\ Chem.\ Acc.\ \textbf{116}, 398 (2006).
%
\bibitem{Hozoi} L.\ Hozoi, U.\ Birkenheuer, P. Fulde, A.\ Mitrushchenkov, and H.\ Stoll, Phys.\ Rev.\ B \textbf{76}, 085109 (2007).
%
\bibitem{FuldeAdv} P.\ Fulde, Adv.\ in Phys.\  \textbf{51}, 909 (2002).
%
\bibitem{GF3} C.\ Buth, U.\ Birkenheuer,  M.\ Albrecht, and P.\ Fulde, Phys.\ Rev.\ B \textbf{72}, 195107 (2005).
%
\bibitem{Gritsenko1} D.\ P.\  Chong, O.\ V.\ Gritsenko, and E.\ J.\ Baerends, J.\ Chem.\ Phys.\ \textbf{116}, 1760 (2002).
%
\bibitem{Gritsenko2} O.\ V.\ Gritsenko and E.\ J.\ Baerends, J.\ Chem.\ Phys.\ \textbf{117}, 9154 (2002).
%
\bibitem{Janak} J.\ F.\ Janak, Phys.\ Rev. B \textbf{18}, 7165 (1978).
%
\bibitem{PerdewIP} J.\ P.\ Perdew, R.\ G.\ Parr, M.\ Levy, and J.\ L.\ Balduz, Jr.\ Phys.\ Rev.\ Lett.\ \textbf{49}, 1691 (1982).
%
\bibitem{GrossRunge} E.\ Runge and E.\ K.\ U.\ Gross, Phys.\ Rev.\ Lett.\ \textbf{52}, 997 (1984).
%
  \bibitem{TDDFT} E.\ K.\ U.\ Gross, J.\ F.\ Dobson, and M. Petersilka, Top.\ Curr.\ Chem.\  \textbf{181}, 81 (1996).
  % 
\bibitem{TDCDFT} G. Vignale amd W. Kohn, Phys.\ Rev.\ Lett.\ \textbf{77}, 2037 (1996);  S.\ K.\ Ghosh and A.\ K.\ Dhara, Phys.\ Rev.\ A \textbf{38}, 1149 (1988).
%
\bibitem{Perdew} J.\ Perdew, Int.\ J.\ Quant.\ Chem.\ \textbf{S19}, 497 (1986).
%
\bibitem{Perdew2}  J.\ P.\ Perdew and S.\ Kurth, \textit{Density Functionals for Non-Relativistic Coulomb Systems: Theory and Applications}, edited by D.\ P.\ Joubert, \textit{Lecture Notes in Physics}, vol. 500, (Springer, Berlin, 1998). 
%
\bibitem{Fuldebook} P.\ Fulde, \textit{Electron Correlations in Molecules and Solids} (Springer-Verlag, Berlin, 1995).
%
\bibitem{FuldeInt} P.\ Fulde, Int.\ J.\ Quant.\ Chem.\  \textbf{76}, 385 (2000).
%
\bibitem{Talman} J.\ D.\ Talman and W.\ F.\ Shadwick, Phys.\ Rev.\ A \textbf{14}, 36 (1976).
%
\bibitem{OEP} T.\ Grabo, T.\ Kreibich, S.\ Kurth, and E.\ K.\ U.\ Gross, \textit{Strong Coulomb correlations in electronic structure: beyond the local density approximation}, edited by V.\ I.\ Anisimov, (Gordon and Breach, London, 1998); R.\ P.\ Muller and M.\ P.\ Desjarlais, J.\ Chem.\ Phys.\ \textbf{125}, 054101 (2006).
%
\bibitem{Stephan} S.\ K\"ummel and J.\ P.\ Perdew, Phys.\ Rev.\ B \textbf{68}, 035103 (2003).
%
\bibitem{EE} M.\ St\"adele, M.\ Moukara, J.\ A.\ Majewski, P.\ Vogl, and A.\ G\"orling, Phys.\ Rev.\ B  \textbf{59}, 10031 (1999).
%
\bibitem{Hedin} L.\ Hedin, Phys.\ Rev.\ \textbf{139}, A796 (1965).
%
\bibitem{Calais} J.\-L.\ Calais, B.\ T.\ Pickup, M.\ Deleuze, and J.\ Delhalle, Eur.\ J.\ Phys.\ \textbf{16}, 179 (1995).
%
\bibitem{Resta1} R.\ Resta, J.\ Phys.\ : Condens.\ Matter \textbf{14}, R625 (2002).
%
\bibitem{Resta2} R.\ Resta, Int.\ J.\ of Quant.\ Chem.\  \textbf{75}, 599 (1999).
%
\bibitem{Resta3} R.\ Resta, Phys.\ Rev.\ Lett.\  \textbf{80}, 1800 (1998).
%
\bibitem{Ladik} J.\ J.\  Ladik, Phys.\ Rep.\  \textbf{313}, 171 (1999).
%
\bibitem{Hirata} S.\ Hirata and S.\ Iwata, J.\ Chem.\ Phys.\ \textbf{109}, 4147 (1998).
%
\bibitem{Malrieu} P.\ Reinhardt and J.\-P.\ Malrieu, J.\ Chem.\ Phys.\  \textbf{109}, 7632 (1998).
%
\bibitem{Marzari} N.\ Marzari and D.\ Vanderbilt, Phys.\ Rev.\ B \textbf{56}, 12847 (1997).
%
\bibitem{Zikovich}  C.\ M.\ Zicovich-Wilson, R.\ Dovesi, and V.\ R.\ Saunders, J.\ Chem.\ Phys. \textbf{115}, 9708 (2001).
%
\bibitem{disent} U.\ Birkenheuer and D.\ Izotov, Phys.\ Rev.\ B \textbf{71}, 125116 (2005).
%
\bibitem{Edmiston} C. Edmiston and K.\ Ruedenberg, Rev.\ Mod.\ Phys.\  \textbf{35}, 457 (1963); J.\ Chem.\ Phys.\ \textbf{43}, S97 (1965).
%
\bibitem{Foster-Boys} J.\ M.\ Foster and S.\ F.\ Boys, Rev.\ Mod.\ Phys.\  \textbf{32}, 300 (1960); S.\ F.\ Boys, Rev.\ Mod.\ Phys.\  \textbf{32}, 296 (1960).
%
\bibitem{Pulaypaos} P.\ Pulay, Chem.\ Phys.\ Lett.\ \textbf{100}, 151 (1983).
%
\bibitem{Horsch1} S.\ Horsch, P.\ Horsch, and P.\ Fulde, Phys.\ Rev.\ B \textbf{28}, 5977 (1983); Phys.\ Rev.\ B \textbf{29}, 1870 (1984).
%
\bibitem{Gr1} J.\ Gr\"afenstein, H.\ Stoll, and P.\ Fulde, Chem.\ Phys. Lett.\ \textbf{215}, 611 (1993).
%
\bibitem{Birken1} V.\ Bezugly and U.\ Birkenheuer, Chem.\ Phys.\ Lett.\ \textbf{399}, 57 (2004).
%
\bibitem{Shukla1} A.\ Shukla, M.\ Dolg, P.\ Fulde, and H.\ Stoll, 
Phys.\ Rev.\ B\ \textbf{60}, 5211 (1999).
%
\bibitem{Abdurahman} A.\ Abdurahman, A.\ Shukla, and M.\ Dolg, J.\ Chem.\ Phys.\  \textbf{112}, 4801 (2000).
%
\bibitem{Stollhoff} G.\ Stollhoff and P.\ Fulde, J.\ Chem.\ Phys.\ \textbf{73}, 4548 (1980).
%
\bibitem{Pulay1} P.\ Pulay and S. Saeb\o, Theor.\ Chim.\ Acta \textbf{69}, 357 (1986).
%
\bibitem{Pulay2} S. Saeb\o, and P.\ Pulay, J.\ Chem.\ Phys.\ \textbf{86}, 914 (1987).
%
\bibitem{Pulay3} S. Saeb\o, and P.\ Pulay, J.\ Chem.\ Phys.\ \textbf{115}, 3975 (2001).
%
\bibitem{Helgaker} See, e.g., T.\ Helgaker, P.\ J\o rgensen, and J.\ Olsen,  \textit{Molecular Electronic-Structure Theory} (John Wiley and Sons Ltd., Chichester, 2000)
%
\bibitem{Stoll2} H.\ Stoll, Chem.\ Phys.\ Lett.\ \textbf{191}, 548 (1992).
%
\bibitem{Paulus} For a review, see B.\ Paulus, Phys.\ Rep.\ \textbf{428}, 1 (2006).
%
\bibitem{Pahl} E.\ Pahl and U.\ Birkenheuer, J.\ Chem.\ Phys.\ \textbf{124}, 214101 (2006).
%
\bibitem{Hirata3} S.\ Hirata and R.\ J.\ Bartlett, J.\ Chem.\ Phys.\ \textbf{112}, 7339 (2000).
%
\bibitem{Pino} R.\ Pino and G.\ E.\ Scuseria, J.\ Chem.\ Phys.\ \textbf{121}, 2553 (2004).
%
\bibitem{GF1} M.\ Albrecht and J.\ Igarashi, J.\ Phys.\ Soc.\ Jpn.\ \textbf{70}, 1035 (2001).
%
\bibitem{GF2} M.\ Albrecht and P.\ Fulde, Phys.\ Status Solidi B, \textbf{234}, 313 (2002).
%
\bibitem{GF4} M.\ Albrecht, Theor.\ Chem.\ Acc.\ \textbf{107}, 71 (2002).
%
\bibitem{Fulde} P.\ Fulde, Theor.\ Chem.\ Acc.\ \textbf{114}, 255 (2005).
%
\bibitem{Cai} Y.\ Cai, L.\ Zhang, Q.\ Zeng, L.\ Cheng, and Y.\ Xu, Sol.\ State Comm.\ \textbf{141}, 262 (2007).
%
\bibitem{Fomichev}V.\ A.\ Fomichev and M.\ A.\ Rumsh, J.\ Phys.\ Chem.\ Solids  \textbf{29}, 1015 (1968).
%
\bibitem{Agui} A.\ Agui, S.\ Shin, M.\ Fujisawa, Y.\ Tezuka, and T.\ Ishii, Phys.\ Rev.\ B \textbf{55}, 2073 (1997).
%
\bibitem{Chrenko} R.\ M.\ Chrenko, Solid State Commun. \textbf{14}, 511 (1974).
%
\bibitem{Renata} R.\ M.\ Wentzcovitch, K.\ J.\ Chang, and M.\ L.\ Cohen, Phys.\ Rev.\ B \textbf{34}, 1071 (1986).
%
\bibitem{Park} K.\ T.\ Park, K.\ Terakura, and N.\ Hamada, J.\ Phys.\ C \textbf{20}, 1241 (1987).
%
\bibitem{Rodrigues} P.\ Rodr\'{\i}guez - Hern\'andez, M.\ Conz\'alez-Diaz, and A.\ Mu\~{n}oz, Phys.\ Rev.\ B, \textbf{51}, 14705 (1995).
%
\bibitem{Douri} Y.\ Al-Douri, Sol.\ State Com.\ \textbf{132}, 465 (2004).

\bibitem{Surh} M.\ P.\ Surh, S.\ G.\ Louie, and M.\ L.\ Cohen, Phys.\ Rev.\ B, \textbf{43}, 9126 (1991). 
%
\bibitem{Lichanot} A.\ Lichanot, P.\ Azavant and U.\ Pietsch, Acta. Cryst. \textbf{B52}, 586 (1996).
%
\bibitem{Orlando} R.\ Orlando, R.\ Dovesi, C.\ Roetti, and V.\ Saunders, J.\ Phys.: Condens. Matter \textbf{2}, 7769 (1990).
%
\bibitem{Kladko} K.\ Kladko and P.\ Fulde, Int.\ J.\ Quant.\ Chem.\ \textbf{66}, 377 (1998).
%
\bibitem{MCSCF} H.\ -J.\  Werner and P.\ J.\ Knowles, J.\ Chem.\ Phys.\ \textbf{82}, 5053 (1985); P.\ J.\ Knowles and H.\ -J.\  Werner, Chem.\ Phys.\ Lett.\ \textbf{115}, 259 (1985). 
%
\bibitem{ICCI} H.\ -J.\  Werner and P.\ J.\ Knowles, J.\ Chem.\ Phys.\ \textbf{89}, 5803 (1988); P.\ J.\ Knowles and H.\ -J.\  Werner, Chem.\ Phys.\ Lett.\ \textbf{145}, 514 (1988). 
%
\bibitem{ICCI2} P.\ J.\ Knowles and H.\ -J.\  Werner, Theor.\ Chim.\ Acta \ \textbf{84}, 95 (1992). 
%
%
\bibitem{Embed} U.\ Birkenheuer, C.\ Willnauer, M.\ von Arnim, W. Alsheimer, and D.\ Izotov, scientific report, Max-Planck-Institut f\"ur Physik komplexer Systeme Dresden, Germany, 2002 (unpublished), Chap. II. 1.8, p. 71 (http://www.pks.mpg.de/mpi-doc/quantumchemistry/report18.html
%
\bibitem{Baranek} P.\ Baranek, C.\ M.\ Zicovich-Wilson, C.\ Roetti, R.\ Orlando, and R.\ Dovesi, Phys.\ Rev.\ B \textbf{64}, 125102 (2001).
%
\bibitem{Hampel} C.\ Hampel and H.\-J.\ Werner, J.\ Chem.\ Phys.\ \textbf{104}, 6286 (1996).
% 
%
\bibitem{Wakatsuki} M.\ Wakatsuki, K.\ Ichinose, and T.\ Aoki, Mater.\ Res.\ Bull.\ \textbf{7}, 999 (1972).
%
\bibitem{Bundy} F.\ P.\ Bundy and R.\ H.\ Wentorf Jr, J.\ Chem.\ Phys.\ \textbf{38}, 1144 (1963).
%
\bibitem{Soma} T.\ Soma, A.\ Sawaoka, and S.\ Saito, Mater.\ Res.\ Bull.\ \textbf{7}, 755 (1974).   
%
\bibitem{Wentorf} R.\ H.\ Wentorf Jr., J.\ Chem.\ Phys.\ \textbf{26}, 956 (1957).
%
\bibitem{Dovesi} R.\ Dovesi, C.\ Pisani, F.\ Ricca, and C.\ Roetti, Phys.\ Rev.\ B \textbf{30}, 972 (1984).
%
\bibitem{Dovesi2} M.\ Caus\`a, R.\ Dovesi, and C.\ Roetti, Phys.\ Rev.\ B \textbf{43}, 11937 (1991).
%
\bibitem{CRYSTAL1} V.\ R.\  Saunders, R.\ Dovesi, C.\ Roetti, M.\ Caus\`a, and N.\ M.\ Harrison\textit{et al.},  CRYSTAL, University of Torino, Italy, 2000.
%
\bibitem{interface} C.\ Roetti, R.\ Dovesi, M.\ von Arnim, W.\ Alsheimer, and U.\ Birkenheuer, the CRYSTAL-MOLPRO  interface, Max-Planck-Institut f\"ur Physik komplexer Systeme, Dresden, Germany, 2002.
%
\bibitem{molpro}
H.\ -J.\  Werner,
P. J. Knowles, 
R. Lindh, 
F. R. Manby, 
M. {Sch\"{u}tz} \textit{et al.}, \textsc{molpro}, Cardiff University, United Kingdom, 2006.
%
\bibitem{Euwema} R.\ N.\ Euwema, G.\ T.\ Surratt, D.\ L.\ Wilhite, and G.\ C.\ Wepfer, Philos.\ Mag.\ \textbf{29}, 1033 (1974).
%
\bibitem{PM} J.\ Pipek and P.\ G.\ Mezey, J.\ Chem.\ Phys.\  \textbf{90}, 4916 (1989).
%
\bibitem{DC} M.\ I.\ Eremets, M.\ Gauthier, A.\ Polian, J.\ C.\ Chervin, and J.\ M.\ Besson, Phys.\ Rev.\ B \textbf{52}, 8854 (1995)
%
\bibitem{Mitrushchenkov} A.\ O.\ Mitrushchenkov and H.\ -J.\ Werner, Mol.\ Phys.\ \textbf{105}, 1239 (2007).
%
\bibitem{Malmqvist} P.\ -\AA.\ Malmqvist, Int.\ J.\ Quant.\ Chem.\ \textbf{30}, 479 (1986).
%
\bibitem{Broer} R.\ Broer and W.\ C.\ Nieuwpoort, Chem.\ Phys.\ \textbf{54}, 291 (1981); Theor.\ Chim.\ Acta, \textbf{73}, 405 (1988).
%
\bibitem{Paper1} A.\ Stoyanova, C.\ Sousa, C.\ de Graaf, and R.\ Broer, Int.\ J.\ Quant.\ Chem.\ \textbf{106}, 2444 (2006).
%
\bibitem{Liviu} L.\ Hozoi, M.\ Laad, and P.\ Fulde, Phys.\ Rev.\ B \textbf{78}, 165107 (2008)
%
\bibitem{SlaterKoster} J.\ C.\ Slater and G.\ F.\ Koster, Phys.\ Rev.\ \textbf{94}, 1498 (1954).
%
 \end{thebibliography}
\end{document}